\documentclass[12pt,preprint]{aastex}
\usepackage{psfig}
\usepackage{emulateapj5,apjfonts,times}
\usepackage[figuresright]{rotating}

\newcommand{\beq}{\begin{equation}}
\newcommand{\eeq}{\end{equation}}
\newcommand{\gapprox}{$\stackrel {>}{_{\sim}}$}
\newcommand{\lapprox}{$\stackrel {<}{_{\sim}}$}
\def\alp{\mbox{$\alpha$}}
\def\farcd{\hbox{$.\mkern-4mu^\circ$}}
\def\farcm{\hbox{$.\mkern-4mu^\prime$}}

\def\arcmin{\hbox{$^\prime$}}

\def\solar{\mbox{$_{\normalsize\odot}$}}

\def\deg{\hbox{$^\circ$}}
\newcommand{\AmS}{{\protect\the\textfont2
  A\kern-.1667em\lower.5ex\hbox{M}\kern-.125emS}}
\newcommand{\lsim}{\ \raise
-2.truept\hbox{\rlap{\hbox{$\sim$}}\raise5.truept\hbox{$<$}\ }}
\newcommand{\gsim}{\ \raise
-2.truept\hbox{\rlap{\hbox{$\sim$}}\raise5.truept\hbox{$>$}\ }}

\slugcomment{}

\righthead{}
\lefthead{D. Gouliermis,\ W. Brandner \& Th. Henning}

\shorttitle{The Initial Mass Function of the association LH 52 in the LMC}
\shortauthors{Gouliermis D., Brandner W. \& Henning Th.}

\begin{document}

\title{The Initial Mass Function toward the low-mass end in the Large 
Magellanic Cloud with HST/WFPC2 Observations} 

%\footnotemark[1]
%\footnotetext[1]{}

\author{D. Gouliermis, W. Brandner, Th. Henning}
\affil{Max-Planck-Institut f\"{u}r Astronomie, K\"{o}nigstuhl
17, D-69117 Heidelberg, Germany}
\email{dgoulier@mpia.de, brandner@mpia.de, henning@mpia.de}
%\author{Gouliermis, D.\altaffilmark{1}}
%\altaffiltext{1}{Max-Planck-Institut f\"{u}r Astronomie, K\"{o}nigstuhl
%17, D-69117 Heidelberg, Germany, dgoulier@mpia.de}

%\date{\today}
%\maketitle
%\centerline{\small\sc Draft Version Printed on \today}

\begin{abstract}
We present $V$ and $I$ equivalent HST/WFPC2 photometry of two areas in the
Large Magellanic Cloud: The southern part of the stellar association LH
52, located on the western edge of the super-shell LMC 4, and a field
between two associations, which is located on the southwestern edge of the
shell, and which accounts for the general background field of the galaxy.
The HST/WFPC2 observations reach magnitudes as faint as $V=25$ mag, much
deeper than have been observed earlier in stellar associations in the LMC.
We determine the MF for main-sequence stars in the areas. Its slope in
both areas is steeper for stars with masses $M$ \lapprox\ 2 M{\solar}
($-$4 \lapprox\ $\Gamma$ \lapprox\ $-$6), compared with stars of $M$
\gapprox\ 2 M{\solar} ($-$1 \lapprox\ $\Gamma$ \lapprox\ $-$2).  Thus, as
far as the field of the LMC concerns the MF does not have a uniform slope
throughout its observed mass range. The MF of the general field of the LMC
was found previously to be steeper than the MF of a stellar association
for massive stars with $M$ \gapprox\ 5 M{\solar}. We conclude that this
seems to be also the case toward lower masses down to $M \sim$ 1
M{\solar}. Our data allow to construct the field-subtracted,
incompleteness-corrected, main-sequence MF of the southwestern part of the
young stellar association LH 52, which accounts for the Initial Mass
Function (IMF) of the system. Its mean slope is found to be comparable,
but still more shallow than a typical Salpeter IMF ($\Gamma \simeq\ -1.12
\pm 0.24$) for masses down to $\sim$ 1 M{\solar}.  We found indications
that the IMF of the association probably is ``top-heavy'', due to the
large number of intermediate-mass stars in the field of the system, while
the general LMC field is found to be responsible for the low-mass
population, with $M$ \lapprox\ 2 M{\solar}, observed in both fields. This
finding suggests that the local conditions seem to favor the formation of
higher mass stars in associations, and not in the background field. No
evidence for flattening of the IMF toward the low-mass regime, nor a lower
mass cutoff in the IMF was detected in our data.
\end{abstract}


\section{Introduction}

Stellar associations are young systems defined by their bright early-type
stellar content (Ambartsumian 1947; Blaauw 1964), constituting a unique
length-scale of the star formation process in a galaxy (Efremov \&
Elmegreen 1998a). While the membership of stars to associations in our own
Galaxy is not easily verified (unless kinematic information is available
for the stars in these areas; e.g. de Zeeuw et al. 1999) extra-galactic
stellar associations are outlined by their bright early-type stellar
content, and hence they can easily be identified in the general stellar
field from the loci of their bright stars alone (Hodge 1986). Thus, their
original identification with the use of photographic or photoelectric data
has been based purely on these stars (see references in Gouliermis et al.
2003, Table 4).  More advanced CCD ground-based observations have allowed
to observe with better sensitivity and to resolve these systems down to
their intermediate-mass populations, in particular for associations in the
Large Magellanic Cloud (e.g. Massey et al. 1989;  Parker \& Garmany 1993;  
Hill et al. 1994; Garmany, Massey \& Parker 1994; Oey \& Massey 1995;  
Will, Bomans \& Dieball 1997;  Parker et al. 2001). A recent comprehensive
study, however, with the high spatial resolution, provided by HST, of
extra-galactic stellar associations and their fields is still lacking.

The Large Magellanic Cloud (LMC), our closest neighboring irregular
galaxy, is characterized by a unique sample of stellar associations (Lucke
\& Hodge 1970) and is known to contain a large number of them throughout
its extent (Gouliermis et al. 2003), easily identified by their bright
stellar content. This galaxy has lower mass and metallicity in comparison
to our own, but it has been shown to have almost the same present-day star
formation rate (e.g. Westerlund 1997). This results in a large number of
early-type stars establishing intense radiation fields, leading to the
star formation in the LMC being different from the Milky Way.  Most of the
LMC associations are related to bright H{\sc II} regions (Davies, Elliott
\& Meaburn 1976), indicating places of on-going star formation. Many of
them have been detected on the edges of super-giant H{\sc I} shells
(Meaburn 1980), as identified by Kim et al. (1999), which suggests that it
is likely that these systems are the product of massive sequential or
triggered star formation events (Gouliermis et al. 2003). They are
characterized by loose structure, so that their observations {\em do not
suffer from crowding effects, while the reddening toward them is low}
enough to reveal the entirety of any pre-main-sequence population. The
stellar content of the general LMC field can be easily identified and
distinguished from the one of the associations, while the foreground
contamination from Galactic stars toward the LMC is minute. These facts
make young LMC stellar associations very attractive targets for the study
of stellar populations in regions of recent star formation and, in
general, for the investigation of newly formed stellar systems in the
Local Group.

%\clearpage
%%%%%%%%%%%%%%%%%%%%%%%%%%%% FIGURE %%%%%%%%%%%%%%%%%%%%%%%%%%%%%%%%%%%%%%%
\begin{figure*}[t!]
\centerline{\hbox{
\psfig{figure=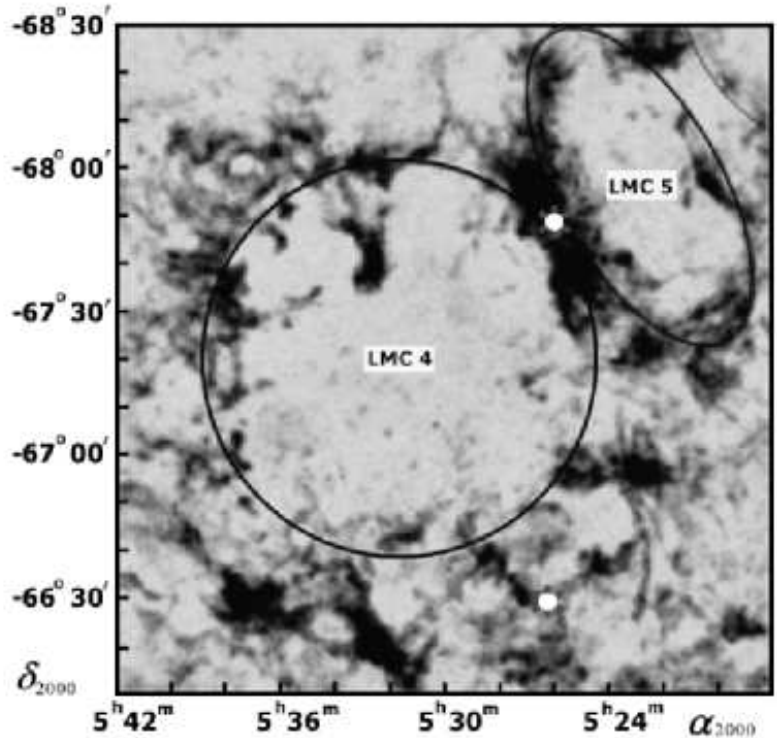,width=11.00truecm,angle=0}
}}
\caption{H{\sc I} map of the area of LMC 4 by Kim et al. (1997).
The two white spots show the loci of the HST/WFPC2 fields studied here:
The southwestern part of the association LH 52 (on the borders of the 
shells LMC 4 and LMC 5) and the ``empty'' field between the associations LH 
54 and LH 55, more than 1{\deg} to the south from the first field.}
\label{lmc4map}
\end{figure*}
%%%%%%%%%%%%%%%%%%%%%%%%%%%%%%%%%%%%%%%%%%%%%%%%%%%%%%%%%%%%%%%%%%%%%%%%%%%
%\clearpage

LMC 4 (Shapley Constellation III) is considered to be one of the most
impressive super-giant shells in the LMC (Meaburn 1980). It is located
north-east of the Bar and its edges, which are easily detected in HI (Fig.
\ref{lmc4map}), are also clearly highlighted by a large number of HII
regions, most of which are related to stellar associations. The origin of
LMC 4 has been discussed by de Boer et al. (1998) and Efremov \& Elmegreen
(1998b). Several previous studies have been concentrated on stellar
associations located on the edge of LMC 4. In the following, we summarize
the results of the three most recent. Dolphin \& Hunter (1998) performed
$UBV$ photometry on six fields within Shapley Constellation III with the
0.9m telescope at CTIO. These fields cover association LH 77 and four LMC
clusters, as well as two general fields. These authors were able to
measure stars accurately down to 19th magnitude in $B$ and $V$, with the
cutoff caused by crowding effects. They concluded that stars have formed
in two modes in the constellation, a more diffuse mode, and a tighter
clustered mode. They constructed the mass function of the observed regions
and they did not find any low-mass cutoff for star formation down to 1.5
M{\solar}. Olsen, Kim \& Buss (2001) analyzed the association LH 72 and
its surroundings in LMC 4 with $UBV$ photometry from 1.5m and 0.9m
telescopes at CTIO. They identified emission stars in the boundaries of
the system and a broad main sequence down to $V \sim$ 21 mag. They
concluded that the expansion of the shell has cleared the northeastern
quadrant of gas and triggered the formation of LH 72 behind its shock
front.  Gouliermis et al. (2002) used $BVR$ and H\alp\ imaging from the
1-m telescope at Siding Spring Observatory to study an area of 20\farcm5
$\times$ 20\farcm5 on the north-east edge of LMC 4, where three stellar
associations are located, the most interesting being the association LH
95. This system was found to be very young, with the highest emissivity of
the related large bright HII region (DEM L 252)  to be connected to four
Be emission stars concentrated in its very center.  Their photometry was
complete down to $V\sim$ 20 mag. These authors constructed the mass
function of the system for masses down to 2 M{\solar}. This limit was set
by the completeness. They suggested that there is a clear distinction
between the population of the system, of its surrounding field and of the
general field of the LMC for stars within the observed stellar mass
limits.

The above studies on associations located in the vicinity of LMC 4
demonstrate the necessity for {\em high spatial resolution photometry} on
the region, in order to detect the lower main-sequence stellar populations
in star-forming regions of the LMC. This will also allow the determination
of the low-metallicity Mass Function (MF) toward the low-mass
end. Since associations are unbound stellar systems, characterized by
loose appearance, contamination by the back- or fore-ground stars of the
LMC itself can be substantial and thus observations of the field of the
galaxy are as important as of the systems themselves. Furthermore, it
should be considered that the observed fields should be wide enough to
cover the extended regions of these systems. Ground-based observations,
with telescopes equipped with Adaptive Optics systems, can provide high
spatial resolution, but require a bright star for wave-front sensing and
are limited to small fields of view. Space observations with HST, on the
other hand, provide the desired resolution with WFPC2 and ACS in a wider
field of view. Therefore, HST can be considered as the best available
instrument for such studies in the optical.

%\clearpage
%%%%%%%%%%%%%%%%%%%%%%%%%%%% FIGURE %%%%%%%%%%%%%%%%%%%%%%%%%%%%%%%%%%%%%%%
\begin{figure*}[t!]
\centerline{\hbox{
\psfig{figure=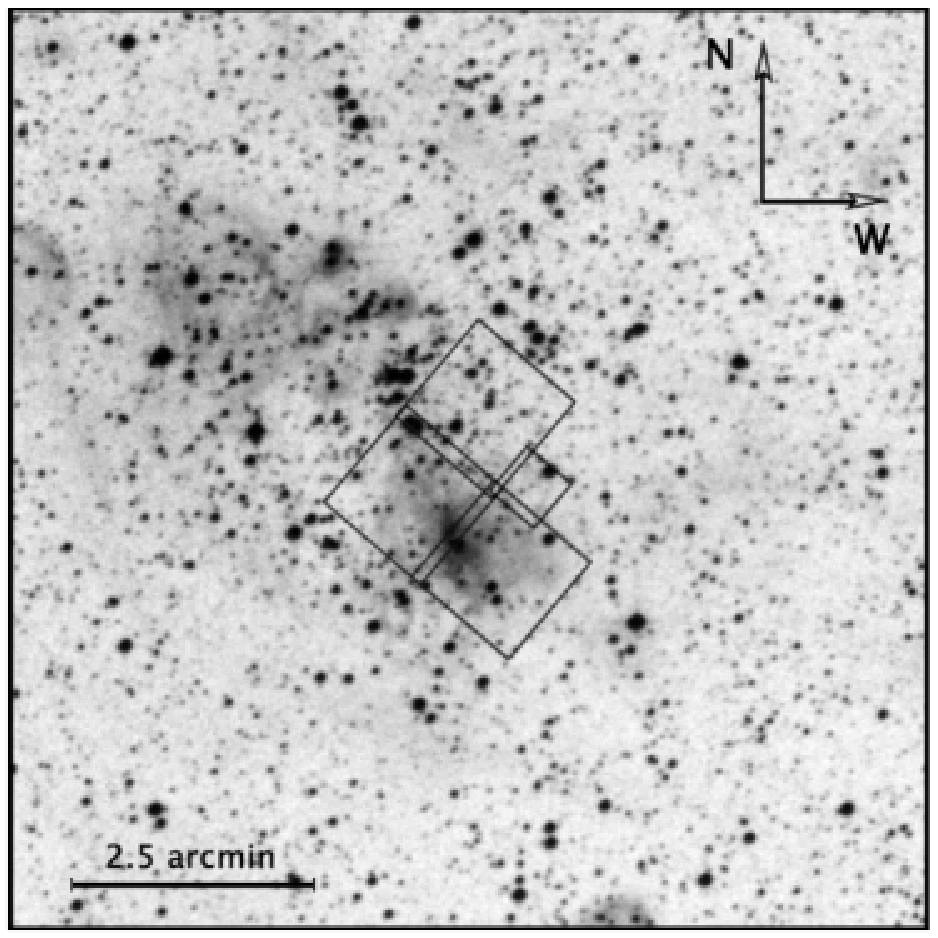,width=8.00truecm,angle=0}
\psfig{figure=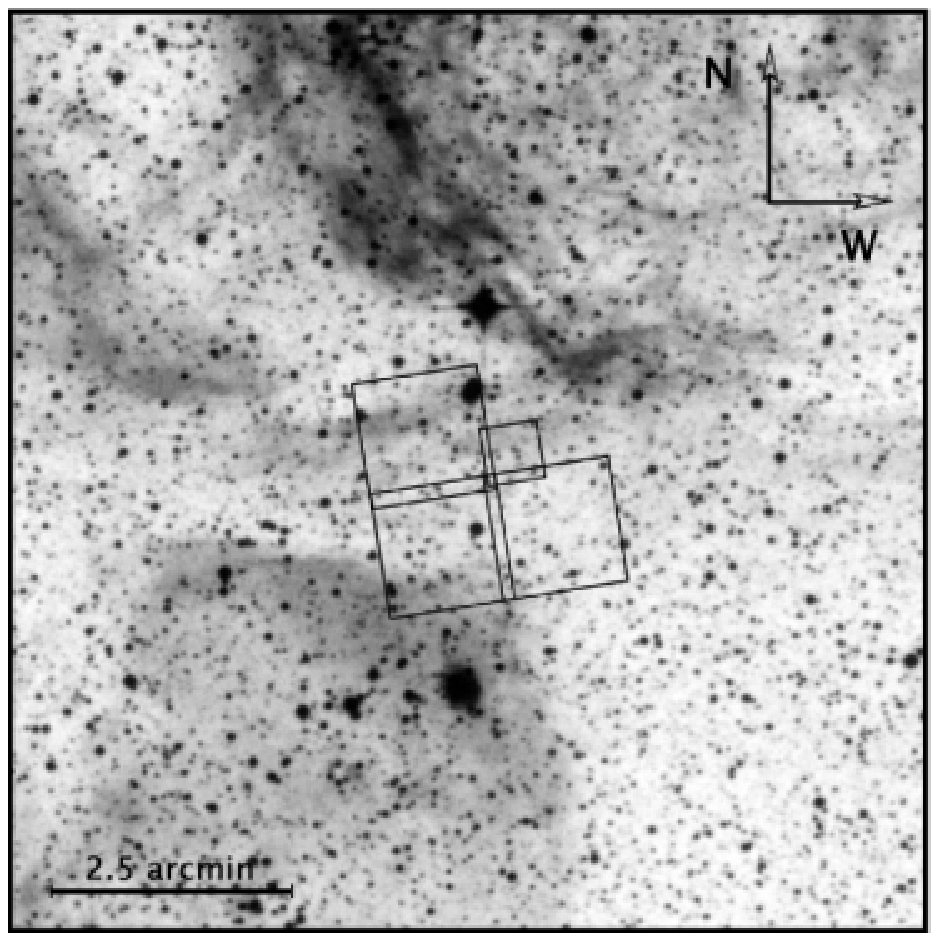,width=8.00truecm,angle=0}
}}
\caption{The HST/WFPC2 pointings on the southern part of the association LH
52 (left) and on the {\em ``field"} (right). The pointings are overlayed on
10\arcmin $\times$ 10\arcmin\ charts of their general areas extracted from
the SuperCOSMOS Sky Survey.}
\label{lh52fov}
\end{figure*}
%%%%%%%%%%%%%%%%%%%%%%%%%%%%%%%%%%%%%%%%%%%%%%%%%%%%%%%%%%%%%%%%%%%%%%%%%%%
%\clearpage

%%%%%%%%%%%%%%%%%%%%%%%%%%%% TABLE 1 %%%%%%%%%%%%%%%%%%%%%%%%%%%%%%%%%%%%%%%%%%
%\clearpage
%%%%%%%%%%%%%%%%%%%%%%%%%%%%%%%%%% TABLE %%%%%%%%%%%%%%%%%%%%%%%%%%%%%%%%%%%%%%%%%%%%%%%%%%
\begin{table*}
%\begin{sidewaystable}
%\begin{center}
\begin{small}
\caption{Log of the observations. Dataset names refer to HST archive catalog. \label{tab1}}
\begin{tabular*}{\textwidth}[]{@{\extracolsep{\fill}}llcccccc}
\tableline
Data Set & Target&       &     &Exposure Time &  R.A.     &Decl.         \\
(name)   & Name  & Filter&Band &(s)           & (J2000.0) &(J2000.0)     \\
\tableline
U5AY120  & LH 52 & F555W &WFPC2 $V$ &1 $\times$ ~~10&05~25~37.66&$-$66~17~02.60 \\
         &       &       &          &2 $\times$ 350&05~25~37.66&$-$66~17~02.60  \\
         &       &       &          &2 $\times$ 350&05~25~37.72&$-$66~17~02.60  \\
         &       & F814W &WFPC2 $I$ &1 $\times$ ~~10&05~25~37.66&$-$66~17~02.60  \\
         &       &       &          &2 $\times$ 350&05~25~37.66&$-$66~17~02.60  \\
         &       &       &          &2 $\times$ 350&05~25~37.72&$-$66~17~02.60  \\
\tableline
U4WOAC0  & LIST-1& F300W &Wide $U$  &1 $\times$ ~~40&05~26~10.42&$-$67~35~07.50 \\
         &       &       &          &1 $\times$ 400&05~26~10.42&$-$67~35~07.50  \\
         &       & F450W &Wide $B$  &1 $\times$ ~~40&05~26~10.42&$-$67~35~07.50 \\
         &       &       &          &1 $\times$ 300&05~26~10.42&$-$67~35~07.50  \\
         &       & F606W &Wide $V$  &1 $\times$ ~~40&05~26~10.42&$-$67~35~07.50  \\
         &       &       &          &1 $\times$ 160&05~26~10.42&$-$67~35~07.50  \\
         &       & F814W &WFPC2 $I$ &1 $\times$ 100&05~26~10.42&$-$67~35~07.50  \\
         &       &       &          &1 $\times$ 400&05~26~10.42&$-$67~35~07.50  \\
\tableline
\tableline
\end{tabular*}
%\end{center}
%\end{sidewaystable}
\end{small}
\end{table*}
%%%%%%%%%%%%%%%%%%%%%%%%%%%%%%%%%%%%%%%%%%%%%%%%%%%%%%%%%%%%%%%%%%%%%%%%%%%%%%%%%%%%%%%%%

%%%%%%%%%%%%%%%%%%%%%%%%%%%%%%%%%%%%%%%%%%%%%%%%%%%%%%%%%%%%%%%%%%%%%%%%%%%%%%%
%\clearpage

Taking under consideration this argument, we use archived HST/WFPC2
imaging data of two selected fields in the LMC, both on the edges of the
super-giant shell LMC 4:  The first covers the southwestern part of a
typical large association LH 52 (Lucke \& Hodge 1970), also known as NGC
1948, and the second, which we will hereon refer to as the {\em ``field"},
covers the area between the associations LH 54 and LH 55 and can be
considered as a typical ``empty" general field in the LMC (Fig.
\ref{lh52fov}). This investigation will provide new information on 1) the
mass function (MF) of the general LMC field and of an association in this
galaxy toward the low-mass end and 2) the faint population of two
fields, one of them covering a {\em bona fide} non-starburst star-forming
region (LH 52).  Deep Color-Magnitude Diagrams (CMD) of both regions, and
consequently their low-mass MF will be presented, while the field
subtracted MF toward the association LH 52 itself will be constructed and
studied with high completeness down to the limit of 1 M{\solar}.

Our main concern in this paper is the study of the MF in the
low-metallicity environment of the LMC. Consequently with the
investigation that follows we add more information on several related
issues: 1) {\em What is the slope of the low-mass end of the MF in this
galaxy?} 2) {\em Is the MF slope universal through the whole detected mass
range?} 3) {\em Is the low-mass MF slope different from the MF slope for
more massive stars?} 4) {\em Is there any specific cut-off in the MF of
young stellar systems in the LMC?} It is worth noting that while previous
studies in the LMC with HST observations concentrated on young star
clusters (e.g. Hunter et al. 1997;  Fischer et al. 1998; Grebel \& Chu
2000; de Grijs et al. 2002; Gouliermis et al. 2004), no information of
such kind is available for stellar associations of this galaxy.

%\clearpage
%%%%%%%%%%%%%%%%%%%%%%%%%%%% FIGURE %%%%%%%%%%%%%%%%%%%%%%%%%%%%%%%%%%%%%%%
\begin{figure*}[t!]
\centerline{\hbox{
\psfig{figure=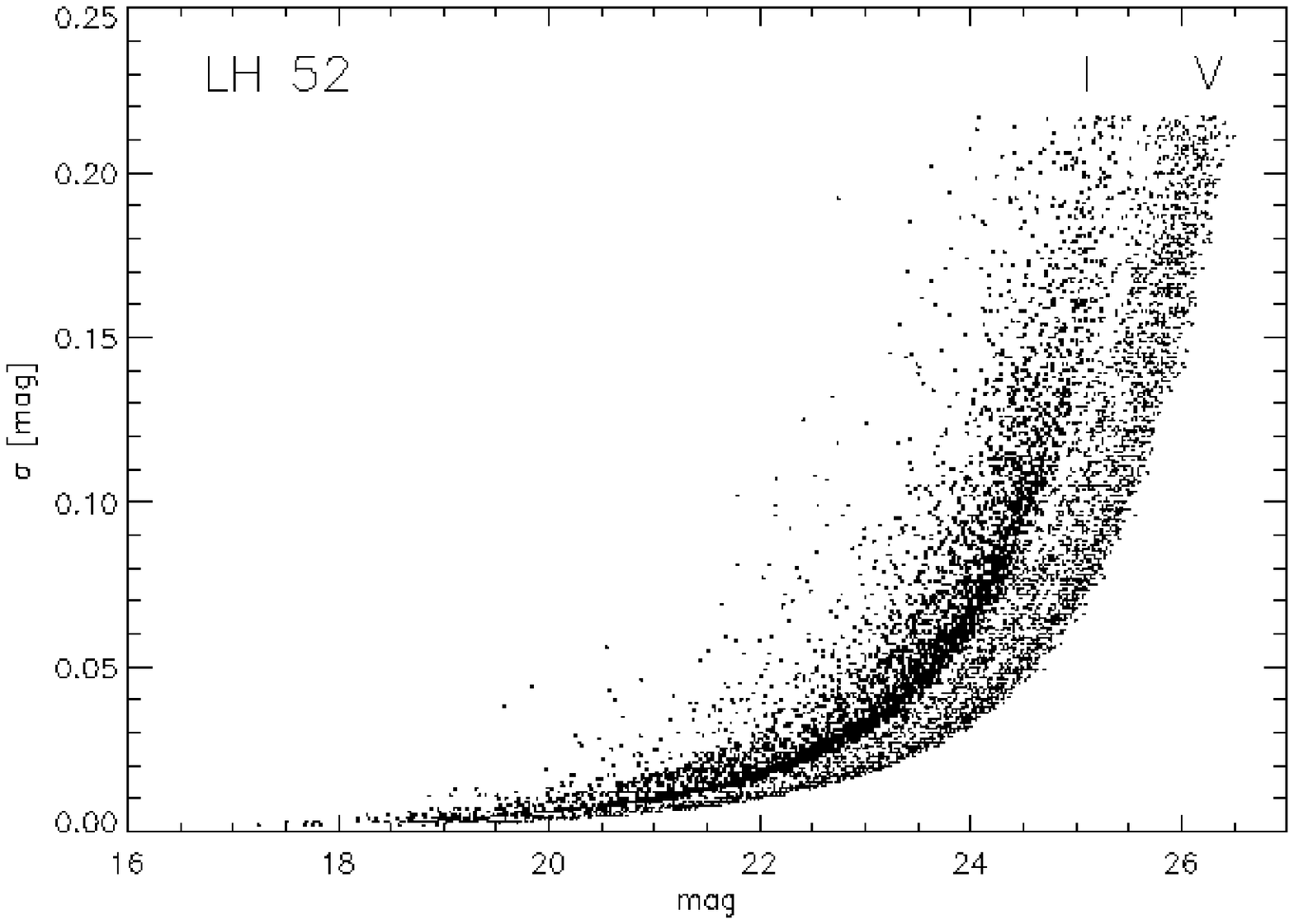,width=8.75truecm,angle=0}
\psfig{figure=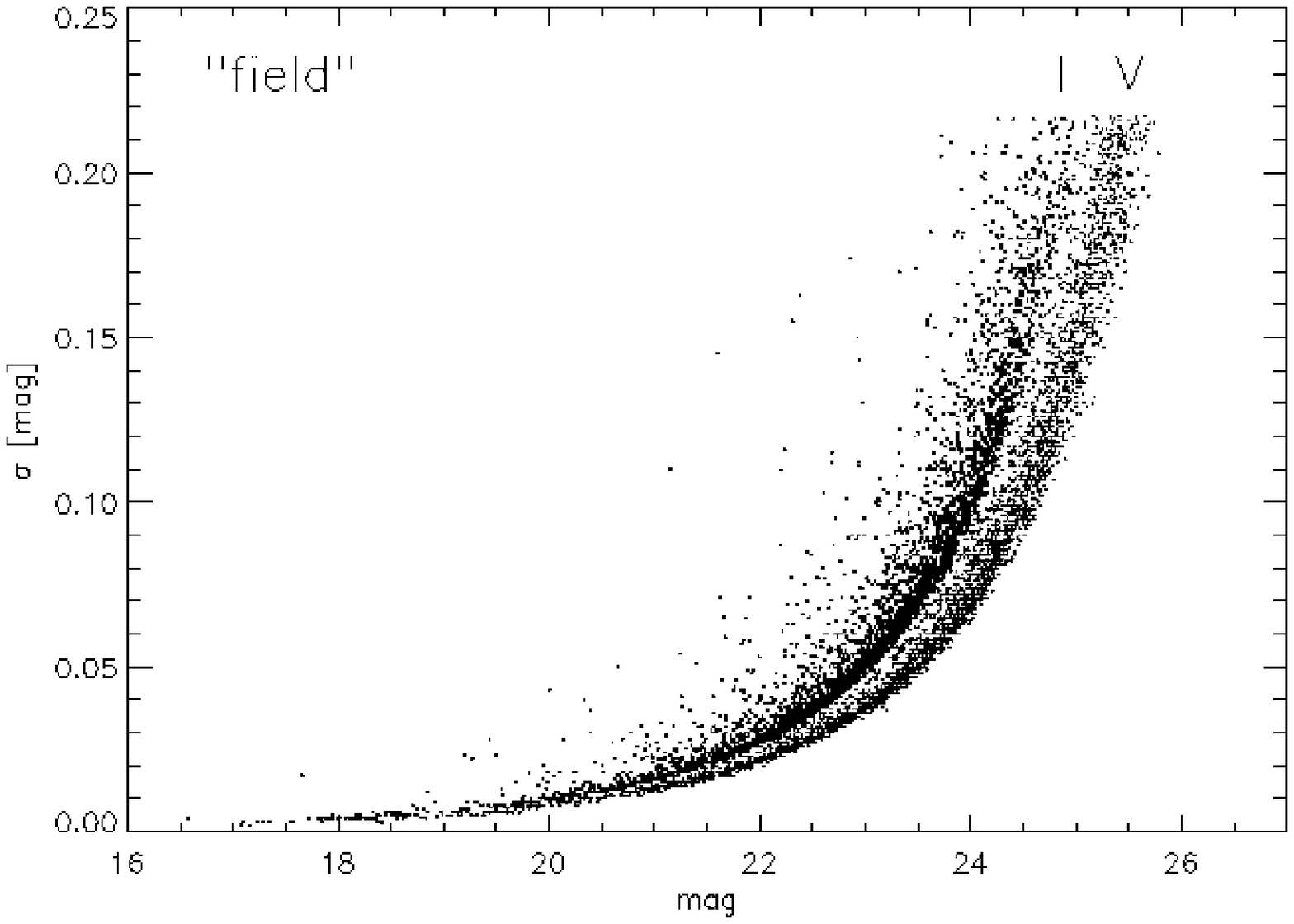,width=8.75truecm,angle=0}
}}
\caption{Uncertainties of photometry as derived by HSTphot
for both $V$ and $I$ bands for the field on LH 52 (left) and the {\em 
``field"} 
(right).}
\label{lh52std}
\end{figure*}
%%%%%%%%%%%%%%%%%%%%%%%%%%%%%%%%%%%%%%%%%%%%%%%%%%%%%%%%%%%%%%%%%%%%%%%%%%%
%\clearpage

The outline of this paper is the following: In the next section (\S 2) we
present our data and we describe the performed photometry. The spatial
distribution of the detected stars in each observed field is investigated
in \S 3, while the investigation of the observed populations takes place
in \S 4, where the color-magnitude diagrams are presented. The
determination and the study of the MF in both fields is described in \S 5,
where we also present the (field subtracted) MF of the observed part of
the association LH 52. The summary of our results and a discussion are
presented in \S 6.

\section{Observations and Data Reduction}

The WFPC2 images on the region of LH 52 were collected as target of the
{\em Hubble Space Telescope} program GO-8134. One telescope pointing was
obtained, with the southern part of the association covered by WF2 and WF3
frames (Fig. \ref{lh52fov} - left panel). The exposure times for the two
filters used {\em F555W} ($\sim V$)  and {\em F814W} ($\sim I$) are given
in Table 1, together with other details of the observations. The data
obtained for the {\em ``field"} covers a larger collection of bands: {\em
F300W} ($\sim U$), {\em F450W} ($\sim B$), {\em F606W} ($\sim V$) and {\em
F814W} ($\sim I$). Details for these observations are also given in Table
1. This area was observed as target of the {\em Hubble Space Telescope}
program GO-8059. Both the WFPC2 fields, overlayed on SuperCOSMOS Sky
Survey images of the general areas, are shown in Fig. \ref{lh52fov}.

The photometry has been performed using the package HSTphot as developed
by Dolphin (2000a).  We used the subroutine {\em coadd} for adding
medium-exposure images (e.g. 2 of 350 sec in each filter in the case of LH
52) to create deep ones. For LH 52 the three sets of exposure times (10,
350, 700 sec) allow us to cover the high dynamic range of brightness of
the stars in the region with good overlap between sets, though we were not
able to detect the brightest ones due to saturation. We recovered these
stars with the use of previous ground-based observations, as it will be
shown later. The fewer exposures used for the {\em ``field"} did not allow
us to create images as deep as for LH 52, and thus the corresponding
stellar sample has a brighter detection limit. The most recent version of
the HSTphot package (version 1.1.5b; May 2003) has been used, which
allows the simultaneous photometry of a set of images obtained by multiple
exposures. HSTphot is tailored to handle the undersampled nature of the
point-spread function (PSF) in WFPC2 images and uses a self-consistent
treatment of the charge transfer efficiency (CTE) and zero-point
photometric calibrations (Dolphin 2000b).

%\clearpage
%%%%%%%%%%%%%%%%%%%%%%%%%%%% FIGURE %%%%%%%%%%%%%%%%%%%%%%%%%%%%%%%%%%%%%%% 
\begin{figure*}[t!]
\centerline{\hbox{ 
\psfig{figure=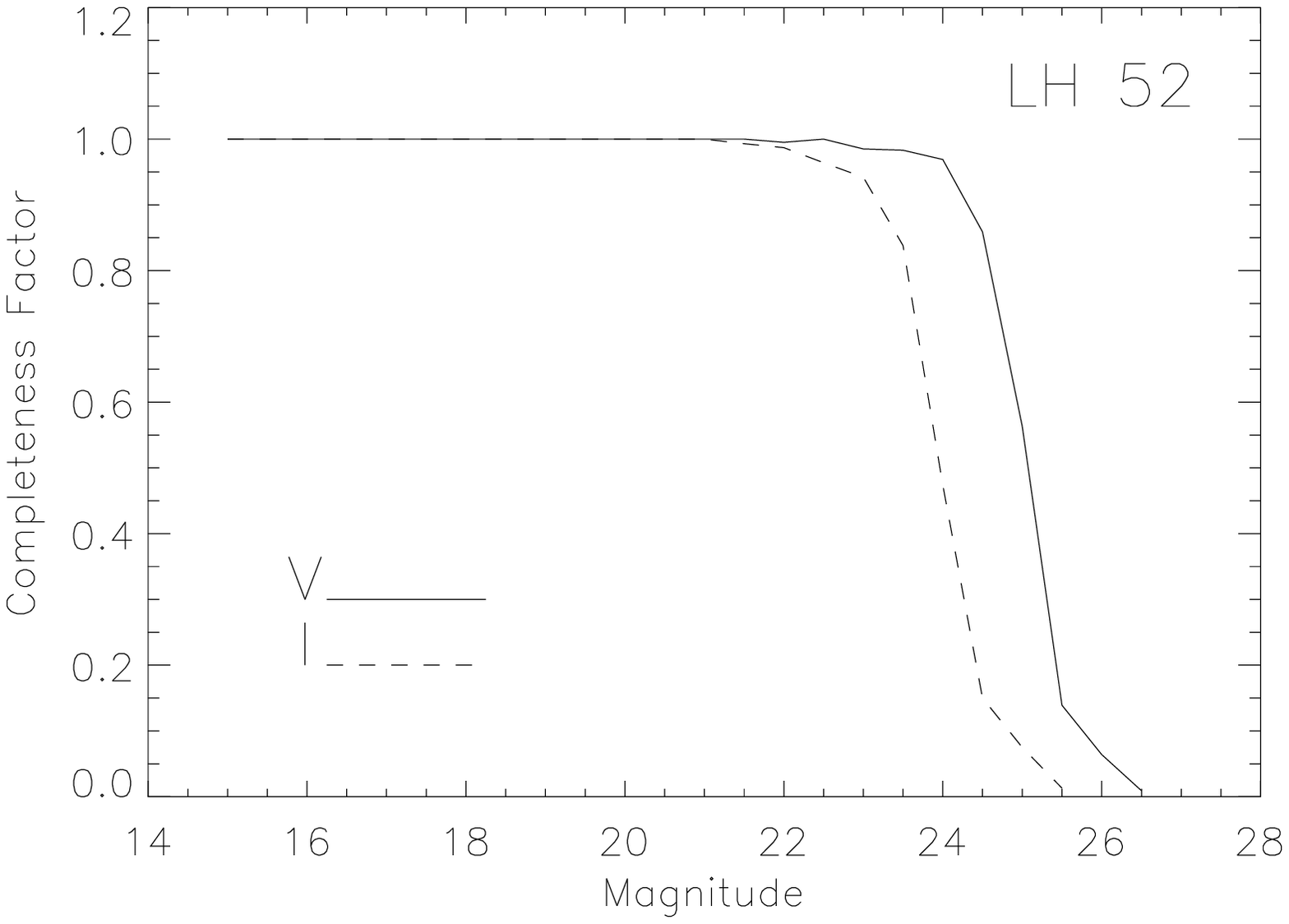,width=8.75truecm,angle=0} 
\psfig{figure=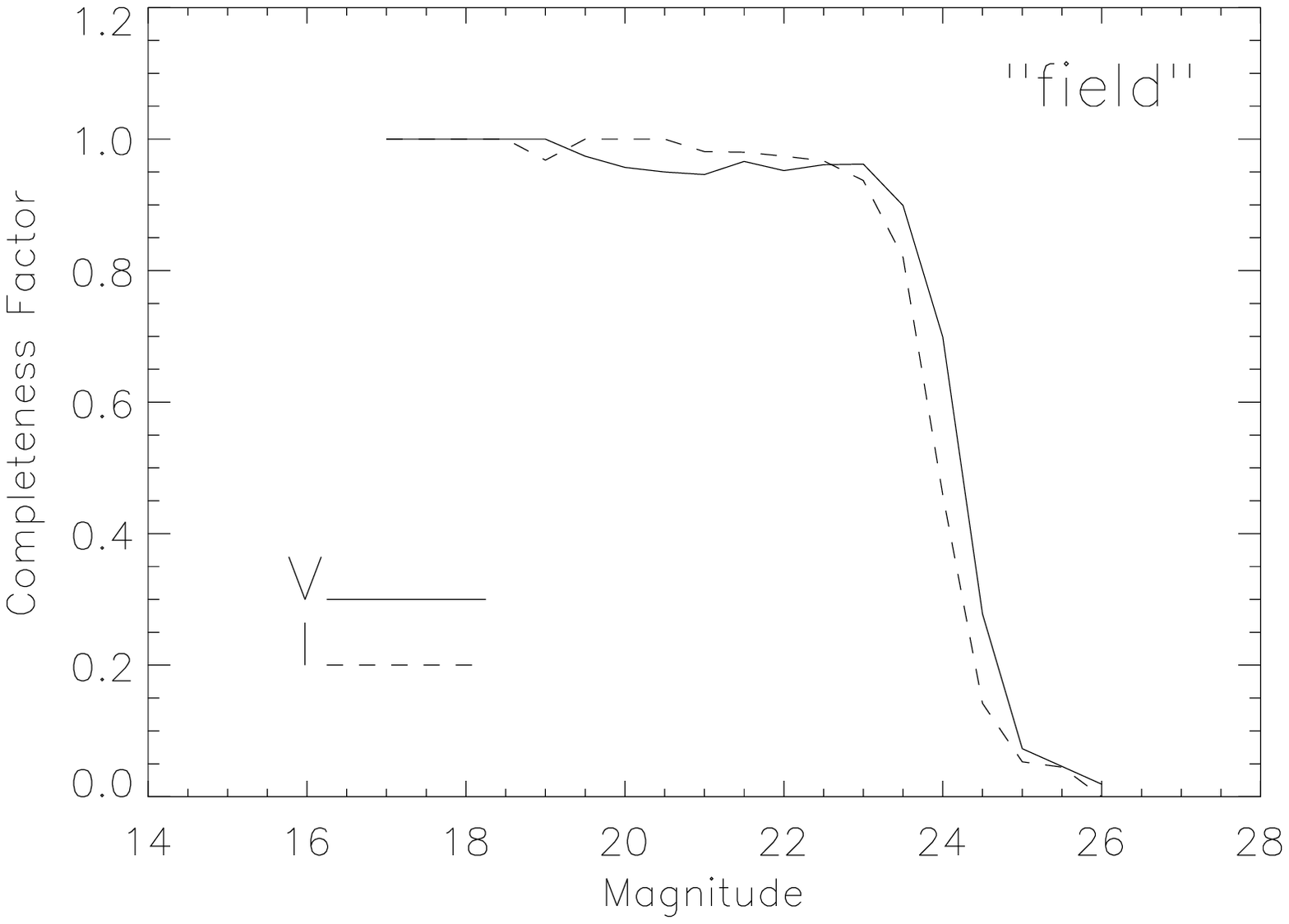,width=8.75truecm,angle=0} 
}}
\caption{Completeness factors as obtained with the observed data set. The
curves obtained for bands $F555W$ and $F814W$ for LH 52 (left) and 
for $F606W$ and $F814W$ for the {\em ``field"} (right).}
\label{lh52cmp} 
\end{figure*}
%%%%%%%%%%%%%%%%%%%%%%%%%%%%%%%%%%%%%%%%%%%%%%%%%%%%%%%%%%%%%%%%%%%%%%%%%%%
%\clearpage

Following Dolphin (2000a) we used the subroutine {\em mask} to take
advantage of the accompanying data quality images by removing bad columns
and pixels, charge traps and saturated pixels.  The same procedure is also
able to properly solve the problem of the vignetted regions at the border
of the chips. The subroutine {\em crmask} was used for the removal of
cosmic rays.  One of the advantages of HSTphot is that it allows the use
of PSFs which are computed directly to reproduce the shape details of star
images as obtained in the different regions of WFPC2.  For this reason, we
adopt the PSF fitting option in the HSTphot routine, rather than use
aperture photometry. The HSTphot photometry routine, {\em hstphot},
returns various data quality parameters for each detected source, which
can be used for the removal of spurious objects. The most useful of them
is the object type, which we used for a first selection of the best
detected stars. According to the recipe outlined in ``{\em HSTphot User's
Guide}'', a good, clean star in an uncrowded field should have a 
goodness-of-fit parameter ($\chi$) of up to 2.5. We used this limit for
our selection. An additional parameter provided by the HSTphot photometry
is the sharpness of each detected object. Reasonable sharpness values for
a good star in an uncrowded field are between $-0.5$ and $0.5$.

CTE corrections and calibrations to the standard $VI$ system were obtained
directly by HSTphot (Dolphin 2000b). Fig. \ref{lh52std} shows typical
uncertainties of photometry as a function of the magnitude for the two
filters, as obtained from the deep-exposure frames. Artificial star
experiments were performed to evaluate the completeness. The HSTPhot
utility {\em hstfake} was used for the creation of simulated images by
distributing artificial stars of known positions and magnitudes. This
utility allows the distribution of stars with similar colors and
magnitudes as in the real CMD. The resulting completeness functions are
shown in Fig. \ref{lh52cmp} for the two selected filters.

%\clearpage
%%%%%%%%%%%%%%%%%%%%%%%%%%%% FIGURE %%%%%%%%%%%%%%%%%%%%%%%%%%%%%%%%%%%%%%%
\begin{figure*}[t!]
\centerline{\hbox{
\psfig{figure=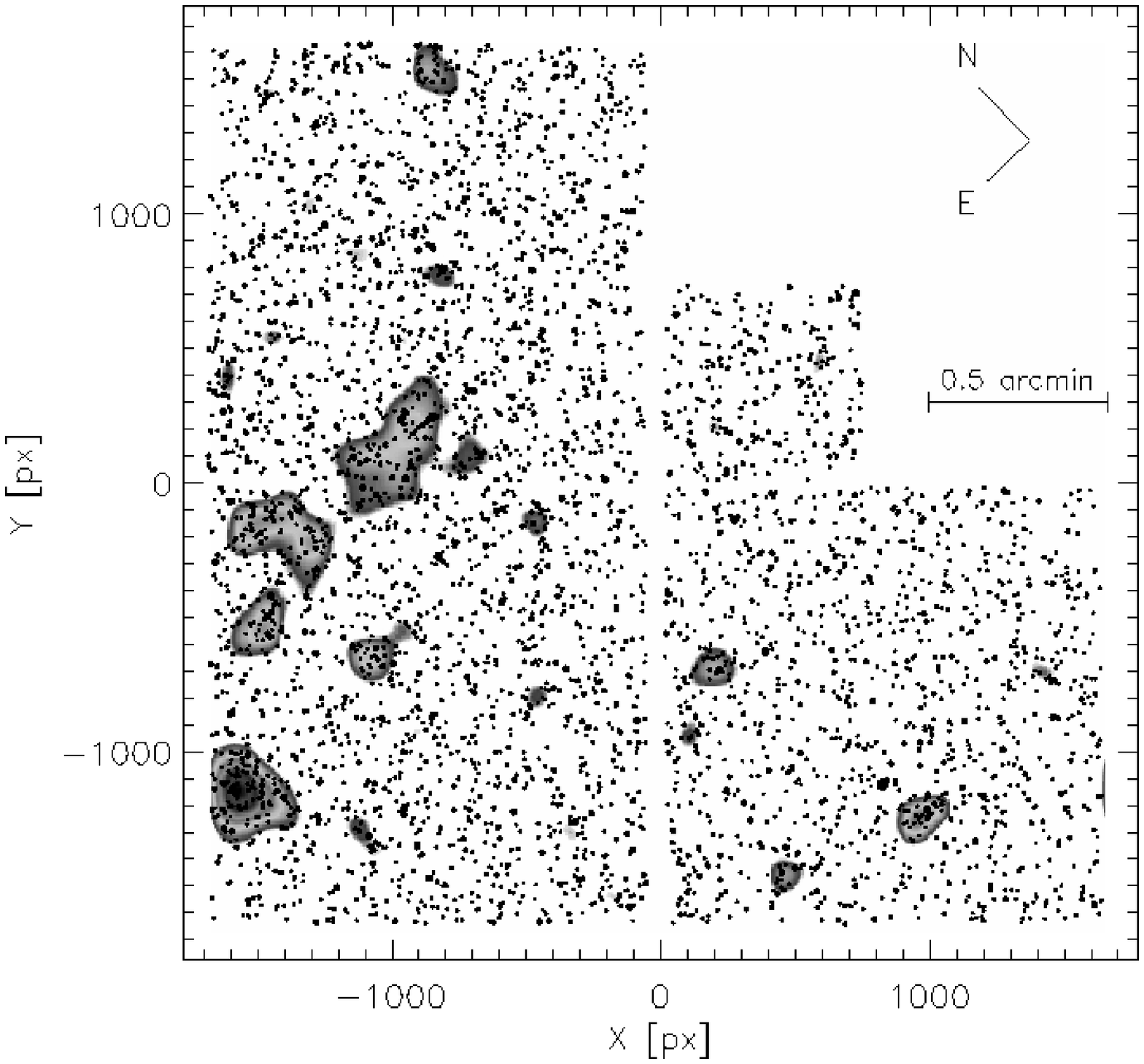,width=8.75truecm,angle=270}
\psfig{figure=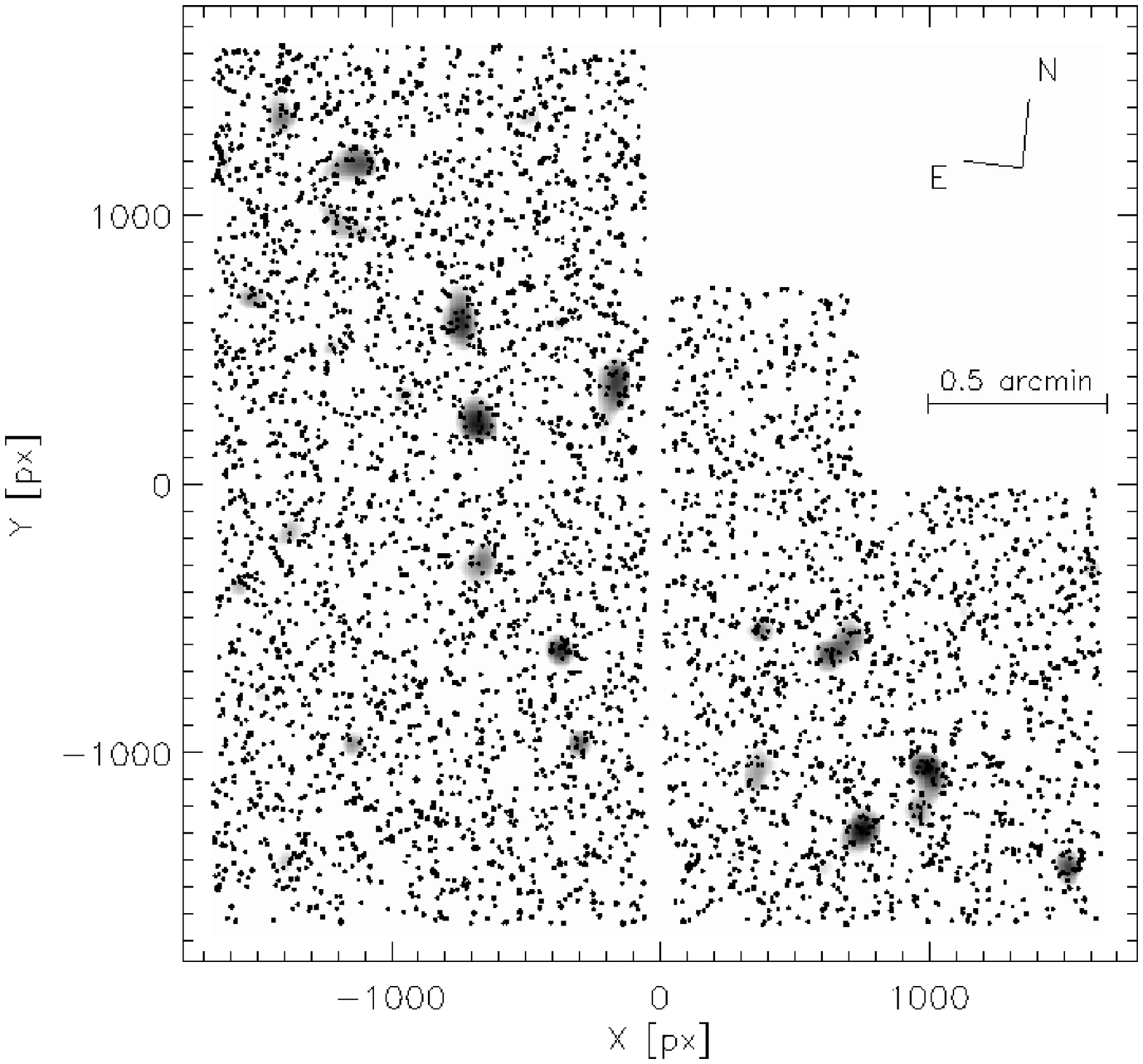,width=8.75truecm,angle=270}
}}
\caption{Star Charts of the observed fields with gray-scale isodensity
contour maps, constructed by star counts of the stellar
catalogues, overplotted for LH 52 (left) and the {\em ``field"} (right). 
The contours start at a level 3$\sigma$ above the local background of each 
field.}
\label{maps}
\end{figure*}
%%%%%%%%%%%%%%%%%%%%%%%%%%%%%%%%%%%%%%%%%%%%%%%%%%%%%%%%%%%%%%%%%%%%%%%%%%%
%\clearpage

\section{Topology of the observed fields}

The HST/WFPC2 field on LH 52 covers the southwestern part of the
association and it is almost entirely within its boundary as it was
defined by Lucke (1972). The boundaries of a stellar concentration can be
defined in a straightforward manner by performing star counts in a
quadrilateral grid of the stars detected in the area. This technique
depends on the size of the covered area, the stellar number statistics and
the selected size of the grid elements, within which stars will be counted
(e.g. Gouliermis et al. 2000). We applied this technique on our data in
order to establish the limits of the stellar concentrations in both
observed WFPC2 fields. We show the star-count images overlayed on the
stellar charts of the HST observations in Fig. \ref{maps}. In the case of
the field on LH 52 we did not use the few stars of the Red Clump (see CMD
in Fig. \ref{lh52cmdiso}), since we are mostly interested in the
distribution of the main-sequence stars. Eventually, the lower MS
population of the CMD can not be distinguished between stellar members of
the association and the field. The star counts on the whole stellar
sample of the field on LH 52 did not show a unique coherent stellar
concentration, as expected for a single stellar system (Fig.~\ref{maps} -
left). This is probably due to the use of local statistics in a relatively
small field such as the one of WFPC2, but it could also indicate
sub-clustering through fragmentation of the parental molecular cloud. In
any case, with star counts we observe the statistically significant local
density peaks, which belong to the main body of the association.  
Specifically, the isodensity map of LH 52 shows that the eastern part of
the observed field is denser populated and large stellar sub-groups, which
belong to the association, are revealed in this region. This
sub-clustering is a phenomenon common among associations and aggregates in
the LMC (Gouliermis et al. 2000).

The clumpy behavior of the area is also detected by star counts on stellar
samples selected according to their brightness.  We divided the stellar
catalog of the area arbitrarily into three groups of stars according to
their magnitudes in $I$. The groups are defined as {\em bright} for stars
with $I$ \lapprox\ 20 mag, {\em intermediate} for those with 20 mag
\lapprox\ $I$ \lapprox\ 23 mag and {\em faint} for the ones with $I$
\gapprox\ 23 mag.  It was found that in some cases the stars, which belong
to different brightness groups have different spatial distributions.  
Specifically, five cases were identified, where clumps of bright stars are
located in areas surrounded by regions with overdensities of faint stars,
meaning that the observed clumps could represent bright newly formed
subclusters. On the other hand, the noted groupings could also be apparent
ones due to the lower completeness of faint stars in the areas, where
bright stars are located, but this does not seem to be the case. We
checked if the change in spatial distribution for the bright clumps
happens mostly close to the 50\% completeness limit ($I \simeq 24$ mag),
or at lower levels. We found that in most of the cases there are larger
numbers of stars with magnitudes of completeness lower than 50\% than with
higher within the boundaries of the bright clumps. This suggests that the
observed segregation between bright and faint stars at the locations of
the bright clumps should not be considered due to incompleteness alone.
The western part of the field on LH 52 is rather empty and it could serve
as a representative field for the local background field of the LMC, due
to its emptiness. Still, there are two factors not allowing us to use it
as such: 1) The fact that the whole field is confined by the boundary of
the association and 2) the argument that the observed emptiness is due to
the use of local statistics within the small field of view, which covers
an area of the association itself, and not due to the lack of
statistically significant stellar numbers.  Consequently, a more careful
treatment should be considered for the use of a general background field
of the LMC. As such serves the second observed area, which we call the
{\em ``field"}.

As shown in Fig. \ref{maps} (right), star counts on the {\em ``field"}
did not reveal any specific stellar concentration, except of few compact
clumps, which are too small to be considered as significant. Since
the area observed in this pointing is fairly small and not very close to
the area of LH 52, there is the issue of how well the {\em ``field"}
represents the general LMC field at the location of LH 52, which should be
addressed. The choice of the {\em ``field"} for characterizing the
general background field of the LMC is supported by the CMD of the area,
which is shown in the next section (Fig. \ref {lh52cmdiso}). There are
several previous investigations on the LMC field observed with HST/WFPC2,
and their results are consistent concerning the star formation history of
the galaxy. In order to check if the use of the {\em ``field"}, as a
representative LMC field is reasonable we compare our results with the
ones of these studies. A field at the outer disk of the LMC was observed
by Gallagher et al. (1996), while Elson, Gilmore \& Santiago (1997)
observed a field in the inner disk of the galaxy. Three fields were
studied by Geha et al. (1998) and seven, more recently, by Castro et al.  
(2001), all of them located at the outer disk. The most recent
investigation is by Smecker-Hane et al. (2002), who studied a field at the
center of the LMC bar and another in the inner disk.

Though these investigations cover a sample of 14 WFPC2 fields, which are
spread in a large area of the LMC (from the center of the bar to the edge
of the outer disk), their color-magnitude diagrams are very similar to
each other and to the CMD of our {\em ``field"}. Specifically, a
prominent red-giant clump at $V-I \simeq$ 1 mag and $V \simeq$ 19 mag,
which characterize these CMDs is also visible in the CMD of the {\em
``field"}. In addition, there is a lack of main-sequence stars brighter
than $V \sim$ 18 mag in all fields. This is not completely due to
saturation, since the LMC field is known to have few massive MS stars per
unit area (Massey et al. 1995). The similarity of the CMDs of all these
WFPC2 fields suggests that the general LMC field does not seem to change
significantly from one area to the other. Consequently, the CMD of the
{\em ``field"} may be used as representative of the LMC general field
population. Still, there are variations in the number of stars observed in
each field. These variations are due to a gradient in the number of stars
in the LMC field, with higher numbers towards the center of the galaxy.
Specifically, each of the pointings observed by Smecker-Hane et al. (2002)
in the bar and the inner disk of the galaxy (1\farcd7 southwest of the
center of the LMC)  covers $\simeq 10^{5}$ stars, in line with $\sim$
15,800 stars observed by Elson, Gilmore \& Santiago (1997) also in the
inner disk. On the other hand Castro et al.  (2001) report that each of
their fields at the outer disk contain, typically, 2,000 stars in
agreement with the number of detected stars by Gallagher et al. (1996) in
their field also in the outer disk. The number of observed stars in the
{\em ``field"} ($\sim$ 4,000) is in line with these variations, since
this field is not located close to the bar, and still it is much closer to
it than all the fields observed in the outer disk. Furthermore, this field
is the closest available to the area of LH 52, and thus it is the most
appropriate to be used as representative of the local LMC field. The facts
presented above give confidence that the choice of the {\em ``field"} as
a representative of the general background LMC population, observed at the
location of LH 52, is reasonable.

\section{Color-Magnitude Diagrams}

\subsection{The area of LH 52 (NGC 1948)}

The $V$ vs. $V-I$ Color-Magnitude diagram (CMD) of the WFPC2 field on LH
52, which includes 4,050 stars is shown in Fig.  \ref{lh52cmdiso}.  Hill
et al. (1995; from hereon HCB95) observed a field including the
associations LH 52 and LH 53 in the LMC with the Ultraviolet Imaging
Telescope (UIT) in the 162 nm bandpass. They also observed two fields
covering the associations from Cerro Tololo Inter-American Observatory
(CTIO) with the 0.9 m telescope in $B$ and $V$. From the 1048 stars
detected by these authors within the boundary of LH 52, 218 are included
in our HST/WFPC2 field, 28 of which were also detected in the far-UV.
HCB95, using 421 stars with 14.8 mag \lapprox\ $V$ \lapprox\ 18.6 mag and
$-$0.20 mag \lapprox\ $B-V$ \lapprox\ $+$0.15 mag, found an average
reddening for LH 52 of $E(B-V)\simeq$ 0.15 with a 1$\sigma$ scatter of
$\pm$0.14 among the individual stars. Isochrone fitting on our CMD showed
that the reddening should be considered to be very low between $E(B-V)=$
0.05 and 0.1 (Fig. \ref{lh52cmdiso}). We adopted a distance modulus
for the LMC of 18.5 $\pm$ 0.1 mag, derived by Panagia et al. (1991) from
SN 1987A.  This corresponds to a distance to the LMC of 50.1 $\pm$ 3.1
kpc. At this distance 1\arcmin\ $\cong$ 14.6 pc. Considering that this is
a star-forming region, the presence of differential reddening in the area
should also be considered.

The comparison of the CMD of the stars detected by HCB95 within the
boundaries of our WFPC2 field with the one presented here for LH 52
exemplifies the gain that has been introduced to crowded stellar
photometry toward LMC with HST. The WFPC2 data, which were taken eight
years later than the ones from CTIO reach a detection limit almost 5
magnitudes deeper. The five brightest of the stars, detected by HCB95
with UIT to have far-UV emission, are main-sequence stars with 13 $<$ $V$
$<$ 15.5 mag and $-$0.2 $<$ $B-V$ $<$ 0.2. They could not be detected in
our observations, due to saturation. These stars show in a direct manner
that the stellar content of this field is very young, possibly still
forming stars. Isochrone fitting on this CMD can only provide an age limit
which coincides with the younger available models of $\sim$ 4 Myr.
Consequently one may suggest that the stellar content of the area is that
young. In Fig. \ref{lh52cmdiso} (left panel) the CMD of the area is
plotted with representative isochrone models of older ages.

\subsection{The area of the ``field"}

\begin{figure*}[t!]
\centerline{\hbox{
\psfig{figure=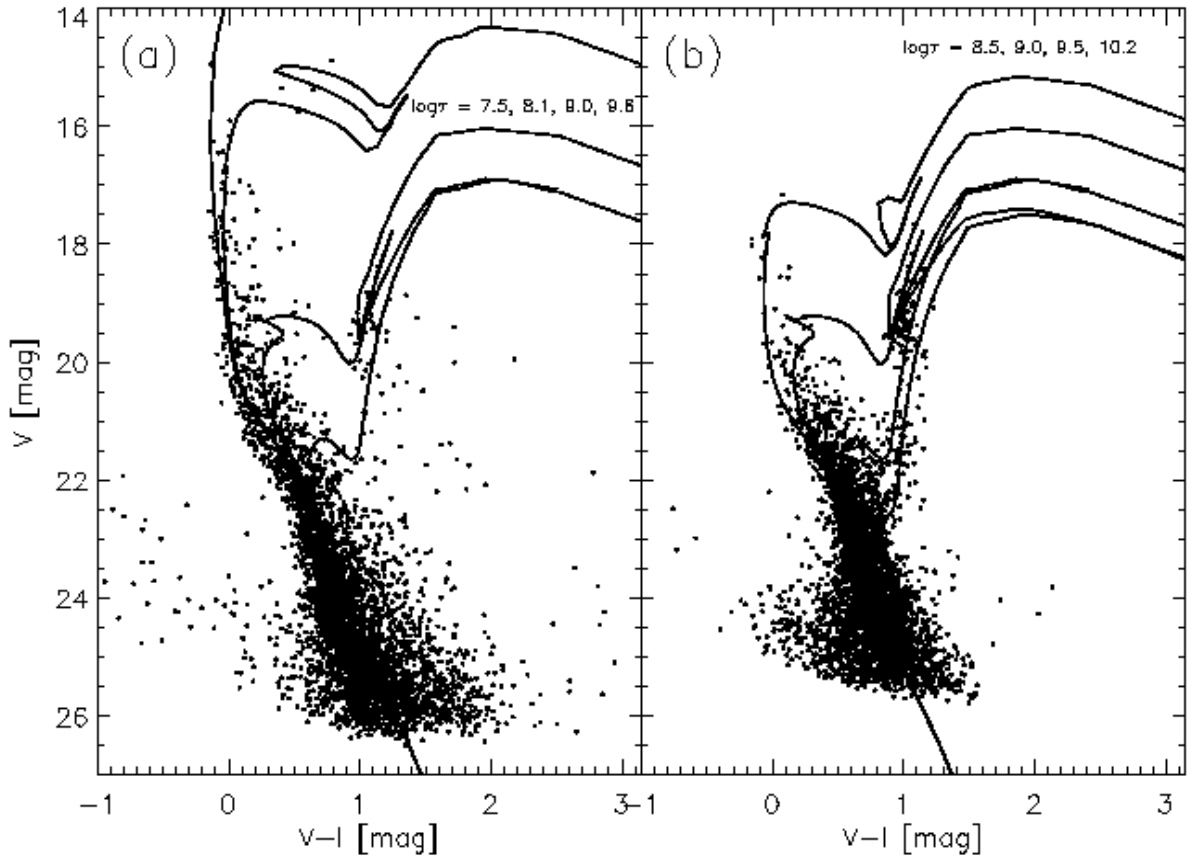,width=17.truecm,angle=0}
}}
\caption{$V-I$, $V$ Color-Magnitude Diagram of the detected stars in both
observed WFPC2 fields: (a) LH 52 and (b) the {\em ``field"}, with
isochrones overplotted.}
\label{lh52cmdiso}
\end{figure*}

The WFPC2 field on LH 52 includes, as it would be expected, also stars of
the general LMC background field, which can be detected in the CMD of Fig.
\ref{lh52cmdiso} (left panel) mostly by the few members of the sub-giant
branch located around $V$ $\simeq$ 20 mag and $(V-I) \simeq$ 1 mag. The
fact that this part of the CMD has only few members suggests that the
contribution of the LMC field in the CMD at these magnitudes of the
association is small. Still, there is a large number of low-mass MS stars
in the field which is expected to contaminate the lower MS of the CMD in
the area of LH 52. In order to quantify this contribution we study the
second HST/WFPC2 field, the {\em ``field"}, which is located on the
``empty'' area almost 1\deg\ south of LH 52 (see Figs. \ref{lmc4map},
\ref{lh52fov}).

The constructed $V$ vs. $V-I$ Color-Magnitude diagram of the {\em
``field"} is shown in Fig. \ref{lh52cmdiso} (right panel). The {\em
``field"} includes 4,155 stars detected in both $V$ and $I$. The
comparison between the two fields does not call for area normalization of
the star numbers, since both cover equal surfaces. One should consider
that differences in stellar numbers between the two fields should be
expected due to a gradient in the LMC field star density, which is higher
closer to the Bar. Indeed, the sub-giant branch of the {\em ``field"} is
very well defined and richer than the one in the CMD of the area of LH 52,
in which the field contamination is very small for $V$ \lapprox\ 18.5 mag.  
The {\em ``field"}, on the other hand, is characterized by lack of upper
main-sequence stars, which has been previously verified for the general
LMC field by other investigators (see \S 3). As far as the lower
main-sequence concerns both CMDs are comparable, except the fact that in
the area of LH 52 there are a lot of reddened low-mass stars, which do not
seem to exist in the area of the {\em ``field"}. This spread of colors
toward the red cannot be accounted for by photometric errors alone. One
cannot exclude the possibility of the existence of pre-main sequence stars
forming this reddened sequence. In the case of the area of SN 1987A, it
was found that such a sequence can be traced very well in the CMD with PMS
models (Panagia et al. 2000).  Differential reddening should also be
considered as important possibility. In the CMDs of Fig. \ref{lh52cmdiso}
is shown that the number of field stars significantly increases towards
fainter magnitudes, and thus the field contamination in the area of LH 52
directly affects the mass function of the lower part of the main sequence.
In addition, according to the tabulation by Ratnatunga \& Bahcall (1985),
we expect negligible contamination by Galactic foreground stars in the
main sequence of both CMDs. Furthermore, according to the work of Metcalfe
et al. (2001)  on the Hubble Deep Fields, the number of background faint
galaxies that may affect the mass functions is expected to be smaller than
the reported uncertainties ($1 \sigma$) in each mass bin for stars
brighter than V $\simeq$ 25 mag.

The reddening toward the area of the {\em ``field"} was estimated with
the use of the available multi-band observations, by constructing the
$BVI$ 2-Color Diagram of this area. We find that the region is
characterized by an absorption of $A_{V}$ \gapprox\ 1 mag. The
corresponding reddening $E(B-V)$ depends on the applied reddening curve
$R_{V}$. Starting with Olson (1975) and Turner (1976), many studies have
been made to establish an accurate value of the ratio $R_{V}$ of the total
absorption in $V$, $A_{V}$, to the color excess, $E(B-V)$:
$A_{V}=R_{V}E(B-V)$. A generally accepted value is $R_{V}=3.1$ (Koornneef
1983). This value is comparable with that found by Mihalas \& Binney
(1981:  $R_{V}=3.2$) and by Leitherer \& Wolf (1984), who found a similar
empirical reddening curve ($R_{V} \simeq 3.13$). We adopt the value
$R_{V}=3.15$ (e.g. Taylor 1986). Consequently the corresponding reddening
toward the region of the {\em ``field"} is found to be $E(B-V)$ \gapprox\
0.3. Isochrone fitting on the CMD (Fig.  \ref{lh52cmdiso}), though, gave a
smaller color excess, comparable to the one toward LH 52, as it was also
found from isochrone fitting.

In order to estimate the field contamination in the area of LH 52 we
rescale the number of stars per magnitude bin in the {\em ``field''}, so
that we can evaluate the number of LMC field stars that have to be
subtracted in each bin of the mass function derived from the original CMD
of LH 52. In order to normalize the stellar numbers per bin in the {\em
``field''} to the corresponding numbers in the field covered by the area
of LH 52, we compare the numbers of stars in several magnitude ranges
for faint main-sequence stars in both CMDs. The most complete stellar
sample (completeness \gapprox\ 50\%) was selected, and the compared
numbers are corrected for incompleteness. We found that the contribution
of the LMC field to the CMD of LH 52 varies from $\sim$ 40\% up to $\sim$
90\% of the background field population observed in the {\em ``field''}.
Specifically, the mean ratio of the number of faint main-sequence stars in
the {\em ``field''} over the one in the area of LH 52 is $0.58 \pm 0.07$,
taking into account the uncertainties due to number counts. This ratio
will be taken under consideration later for the appropriate normalization
of the numbers of stars per mass bin in the MF of the {\em ``field''},
which will be used for the subtraction of the background field from the MF
of the area of LH 52. It is worth noting that, since the sub-giant branch
seen in both the CMD of the {\em ``´field´´} and the one of LH 52
characterize mostly the background population, we made the same comparison
for the numbers of stars in the sub-giant branch. We found that the ratio
of the number of these stars in the {\em ``field''} over the one in the
area of LH 52 is $0.48 \pm 0.10$, comparable to the one given above. In
summary, the MF presented in the following section has been corrected for
completeness and contamination by LMC field stars.  We have investigated
possible contamination by galactic field stars and faint galaxies, and in
both cases found their contributions to be negligible.

%\clearpage
%%%%%%%%%%%%%%%%%%%%%%%%%%%% FIGURE %%%%%%%%%%%%%%%%%%%%%%%%%%%%%%%%%%%%%%%
%%%%%%%%%%%%%%%%%%%%%%%%%%%%%%%%%%%%%%%%%%%%%%%%%%%%%%%%%%%%%%%%%%%%%%%%%%%
%\clearpage

\subsection{Stellar populations in the observed fields}

Previous HST studies on the population of the LMC field have investigated
the star formation history of the galaxy. Consequently, our results of the
study of the {\em ``field''} may complement the previous ones. Gallagher
et al. (1996) used WFPC2 to observe a patch of the outer disk of the LMC.
They constructed the $V$- and $I$-equivalents CMD for more than 2,000
stars down to $V \sim$ 25 mag and they used stellar population models to
constrain the star formation history within the past 3 Gyr.  The
distribution of sub-giant stars indicate that a pronounced peak in the
star formation rate occurred in this region about 2 Gyr ago. If lower
main-sequence stars in the LMC have moderate metallicities, then the age
for most LMC disk stars is less than about 8 Gyr. Elson, Gilmore \&
Santiago (1997) presented $VI$ HST/WFPC2 photometry, which extends down to
$V \sim$ 25 mag, for a region in the inner disk of the LMC and they
confirmed the previous result indicating that an intense star formation
event in the disk of the galaxy occurred a few times 10$^{9}$ yrs ago. A
more recent study by Castro et al. (2001)  focused on seven fields $\sim$
5\deg\ away from the LMC center, observed also in $V$- and $I$-equivalents
with WFPC2. Their deep photometry ($V$ \lapprox\ 26 mag) and the use of
isochrones showed that an old population ($>$ 10 Gyr) exists in all
fields. These authors verified the presence of enhanced star formation,
which took place 2 to 4 Gyr ago, in the fields localized in the north to
north-west regions. The CMD of the {\em ``field''} shows indications that
verify these results. In Fig.  \ref{lh52cmdiso} (right panel) we present
this CMD with four indicative isochrone models overplotted. The models of
the Padova group in the HST/WFPC2 magnitude system (Girardi et al. 2002)
were used. From the isochrone fits in Fig. \ref{lh52cmdiso} is indeed
shown that this field is characterized by a significant star formation
event that took place 1 - 3 Gyr ago. In addition, the sub-giant region of
the CMD is very well traced by models of older age ($\sim$ 15 Gyr).

The CMD of LH 52 area gives the impression that this is an area of mixed
populations. This is expected since the field on LH 52 would not only
include a star-forming association, but also a contribution of the general
LMC field. There are several facts, which favor the conclusion that the
observed part of the association LH 52, which is covered by our first
WFPC2 field, is a star-forming region. It is related to a HII region named
DEM L 189a (Davies, Elliot \& Meaburn 1976) or N 48B (Henize 1956), which
seems to coincide with the small diffuse nebula seen in Fig. \ref{lh52fov}
(left)  on the limit between WF3 and WF4 frames of the WFPC2. From the
observations presented here, it can be seen that this nebula is possibly
excited by one single bright star, on which it is centered. The general
area of the association is characterized by several bright nebulae. Will
et al. (1996) performed UV and optical spectroscopy on bright blue stars
located in this area and they derived an age of 5 - 10 Myr.  They
classified seven stars as B-type and they found three O stars. All of
those stars are located outside our WFPC2 field, with the O-stars being
far away. One X-ray source was identified by ROSAT in the area
covered by our WFPC2 field, which is also characterized by the presence of
one MSX (8 $\mu$m) and an IRAS source. Furthermore, a CO over-density
found with NANTEN (Yamagughi et al. 2001) located to the northwest of our
field. Hence, taking under consideration also the existence of OB stars
found by Will et al. (1996) and of bright upper MS UV-emitters detected by
HCB95, one can conclude that possibly the whole area of LH 52, and
consequently the observed region, is in fact a location, where star
formation is taking place. Under these circumstances one cannot exclude
the possibility of the existence of pre-main sequence (PMS) stars in our
CMD, since the reddening toward this region is low, so that faint and cool
objects could be easily detected in the optical.

Several studies in our galaxy have revealed PMS populations in the regions
of stellar associations. Specifically, Massey, Johnson \& Degioia-Eastwood
(1995) argue that the regions of the OB associations of the Galaxy show
evidence of PMS stars with masses between 5 - 10 M{\solar} and ages $\tau$
$<$ 1 Myr, while 95 low-mass PMS stars with an age of $\sim$ 5 Myr were
discovered in the Upper Scorpius OB association by Preibisch \& Zinnecker
(1999). The young stellar content of the galactic starburst NGC 3603 has
been investigated with ground-based NIR observations by Brandl et al.
(1999) and more recently Stolte et al. (2004) who detected PMS stars with
masses down to 0.1 M{\solar} and an age in the core region of NGC 3603
between 0.3 - 1 Myr. Similar studies in the LMC concentrated on the
starburst region of 30 Doradus. Brandl et al. (1996) performed adaptive
optics NIR observations of the central region of 30 Doradus (R 136) and
they detected 108 ``extremely red sources'', which are most likely PMS
stars of low or intermediate mass.  A red population with masses down to
1.35 M{\solar}, well traced by PMS isochrones, was also discovered in the
CMD of R 136 by Sirianni et al. (2000) with the use of HST/WFPC2 data,
though no variable extinction was taken into account (Andersen 2004). In
addition, Brandner et al. (2001) observed 30 Doradus with HST/NICMOS and
they found no evidence for a lower mass cutoff for pre-main sequence stars
in the area.  Concerning the LMC field, Panagia et al. (2000)  have
identified several hundreds pre-main-sequence stars through their H{\alp}
emission in the field around Supernova 1987A and they found that most of
them have masses in the range 1 - 2 M{\solar} and ages between 1 - 2 and
20 Myr. These authors suggest that some star formation is still ongoing in
this field.

Similar investigations on stellar associations in the LMC are still
lacking. The low reddening toward their regions and the small crowding in
their areas imply that {\em LMC associations should be considered among
the best tracers of extra-galactic PMS populations}. We used the grid of
bolometric corrections presented by Keller, Bessell \& Da Costa (2000) for
populations in the LMC to construct the H-R diagrams (HRD) of the general
stellar populations found in both WFPC2 fields presented here. It is
interesting to note that in the diagram of the LH 52 area we could trace
faint red sources with PMS isochrone models for 2.5 - 20 Myr from Siess et
al. (2000). Still, this cannot be considered as a detection of PMS stars
in regions of associations. Since PMS stars occupy a different place
in the NIR CMD than MS stars, they are expected to be easily
distinguishable in LMC associations from their background field. Hence,
high-resolution NIR observations toward such regions are needed for the
definite detection of PMS stars and the construction of their MF.
Consequently for the construction of the optical MF here, we treat all the
detected faint stars in both fields as evolved MS stars.

\section{Mass Functions of the System and of the Field}
\subsection{Construction of the IMF}

The distribution of stellar masses formed in a given volume of space in a
stellar system is known as the Present Day Mass Function (PDMF) of the
system. The Initial Mass Function (IMF) of the system can then be
constructed, assuming that all stars were born simultaneously, after
disentangling the evolutionary effects of the stars. There are various
parameterizations of the IMF (see e.g. Kroupa 2002). A commonly used
parameterization is the one proposed by Scalo (1986), where the IMF is
characterized by the logarithmic derivative $\Gamma$, called index: $$
\Gamma = \frac{{\rm d}\log{\xi(\log{\rm m})}}{{\rm d}\log{\rm m}} $$ Here
$\xi(\log{\rm m})$ is the IMF and $\Gamma$ is its slope. It can be derived
from the linear relation of $\log{\xi(\log{\rm m})}$ and $\log{\rm m}$.
Reference value for the IMF slope is index $\Gamma$ as found by Salpeter
(1955) for the solar neighborhood ($\Gamma = -1.35$, for a mass range 0.4
\lapprox\ $m$/M{\solar} \lapprox\ 10). For measuring this quantity for a
single stellar system one can assume that all stars in the system are the
product of star formation, which spans over time-scales of few
Myrs. This is more or less true, especially for young systems like stellar 
associations. Thus, assuming that all stars in an association were born 
within the last 10 Myr (which should be the case for LH 52) one can refer 
to the system's MF as its IMF.

The MF of a stellar system is constructed by counting stars in mass
intervals. This can be achieved by two methods: (1) By translating their
luminosities into masses using mass-luminosity relations (e.g. de Grijs et
al. 2002) and then constructing the distribution of the derived masses and
(2) by directly counting stars between evolutionary tracks according to
their positions in the HRD (e.g. Massey et al. 1995). It is almost certain
that the second method is more straightforward, because no transformation
of luminosities to masses is required, while for the first method there is
a definite dependence on the isochrone models used for the transformation
of luminosities to masses. Still, this dependence results only to a 
systematic offset of the overall masses in comparison to the second method 
(de Grijs et al. 2002). Gouliermis et al.  (2004) presented their
results on the mass segregation effect in four Magellanic Clouds clusters
and their mass functions and they argue that there is also a dependence on
the theoretical models for the second method. These authors compared the
MF slopes derived with both methods for their clusters and they found that
in general both counting methods seem equally adequate for the
construction of the MF giving comparable results (their figure 7).

We verified this result by constructing the MF of the LH 52 area with both
methods. For the first method the MF was determined adopting a
mass-luminosity relation derived from theoretical evolutionary models. We
used the latest Padova theoretical isochrones in the HST/WFPC2 STmag
system (Girardi et al. 2002). These models were originally developed by
Girardi et al. (2000)  and they were transformed to the HST/WFPC2
pass-bands as described in Salasnich et al. (2000). For the second method,
transformation of our measured $V-I$ colors and $V$ magnitudes into
temperatures and luminosities was made through interpolation into a grid
of synthetically derived colors and bolometric corrections presented by
Keller, Bessell \& Da Costa (2000). Preliminary results on the second
method have been already presented by Gouliermis, Brandner \& Henning
(2004). The evolutionary tracks presented by Schaerer et al. (1993) were
used for counting stars in the HRD. An excellent agreement on the main
sequence between these tracks from Geneva and earlier isochrones from
Padova (Fagotto et al. 1994) has been found. In all models the LMC
metallicity (Z=0.008) was taken into account.  The counted stellar numbers
were corrected for incompleteness and were normalized to a surface of 1
kpc$^{2}$ for both methods. We computed the slope of the MF constructed
with each method and we found that it does not differ considerably, being
the same within the errors, for the same mass ranges.  This coincidence
gives confidence that the use of only one method for constructing the IMF
will be sufficient.

%\clearpage
%%%%%%%%%%%%%%%%%%%%%%%%%%%% FIGURE %%%%%%%%%%%%%%%%%%%%%%%%%%%%%%%%%%%%%%%
\begin{figure*}[t!]
\centerline{\hbox{
\psfig{figure=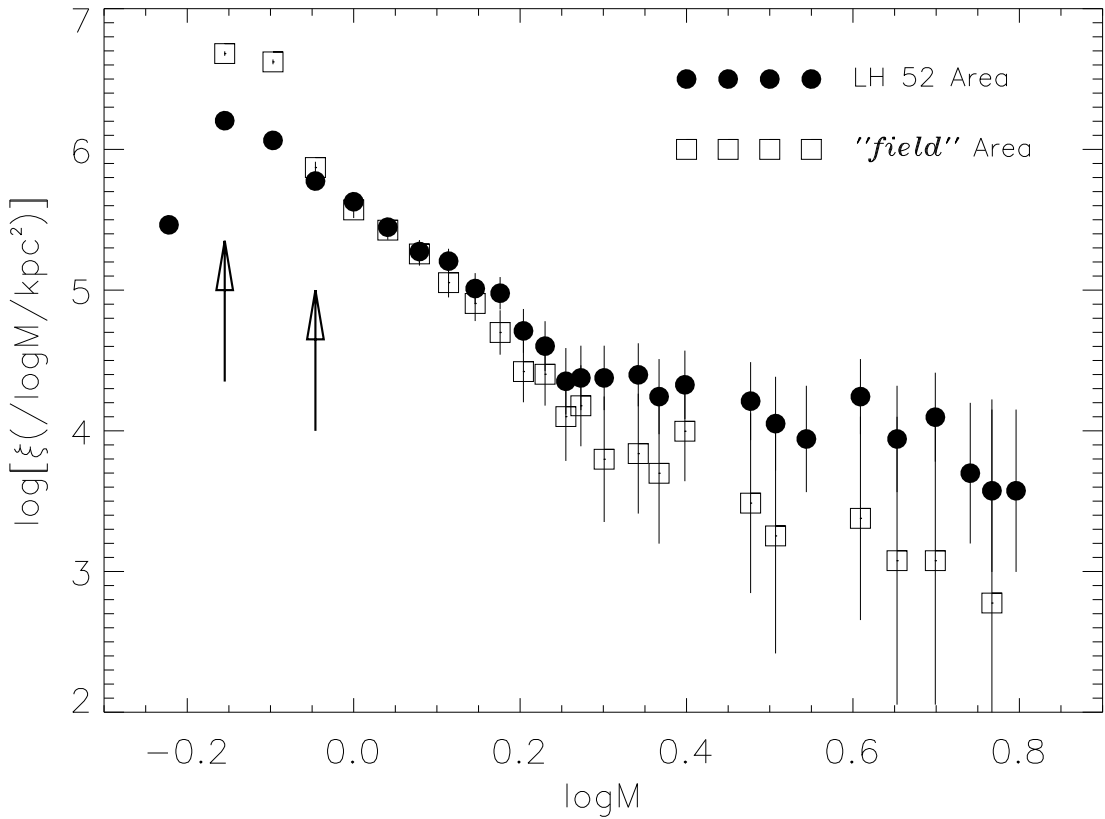,width=17.truecm,angle=0}
}}
\caption{The main-sequence MF of the LH 52 area and the {\em ``field''}
constructed by using the mass-luminosity relation provided by the
theoretical models. The stars have been counted in logarithmic (base ten)
mass intervals. The numbers have been corrected for incompleteness and
normalized to a surface of 1 kpc$^{2}$.  The errors reflect the Poisson
statistics. The arrows indicate the 50\% completeness level for the LH 52 
area (left arrow) and the {\em ``field''} (right arrow).}
\label{imfs}
\end{figure*}
%%%%%%%%%%%%%%%%%%%%%%%%%%%%%%%%%%%%%%%%%%%%%%%%%%%%%%%%%%%%%%%%%%%%%%%%%%%
%\clearpage

\subsection{The Mass Function of the observed fields}

We derived the MF of the MS stars in the LH 52 area and the {\em
``field''}, by selecting them from the corresponding stellar samples and
using the mass-luminosity relation provided by the theoretical models of
Girardi et al. (2002) for the determination of their initial masses. The
stars were counted in logarithmic (base ten) mass intervals. The counted
numbers were corrected for incompleteness and normalized to a surface of 1
kpc$^{2}$. The MF of both areas is given in Fig. \ref{imfs}. The errors
reflect the Poisson statistics of the counting process and they are
suitably corrected and normalized. In this figure it is shown that the MF
slope toward the low-mass end seems to be different from the one toward
the intermediate-mass range, with the MF being steeper toward the lower
masses. The limit of $M \simeq 2$ M{\solar} may be considered as a
reasonable threshold separating the two mass ranges having different
slopes. This phenomenon of different MF slopes between mass ranges with
$M$ \gapprox\ 2 M{\solar} and $M$ \lapprox\ 2 M{\solar} is more prominent
in the case of LH 52 area, while the MF of the {\em ``field''}, as seen in
Fig. \ref{imfs}, could be considered as a single slope distribution.
Considering that each observed area represents a region of different
context in the LMC (the first being a star-forming region and the second
an empty field), these results support the suggestion of a top heavy MF
for typical star-forming regions in the LMC.

\begin{figure*}[t!]
\centerline{\hbox{
\psfig{figure=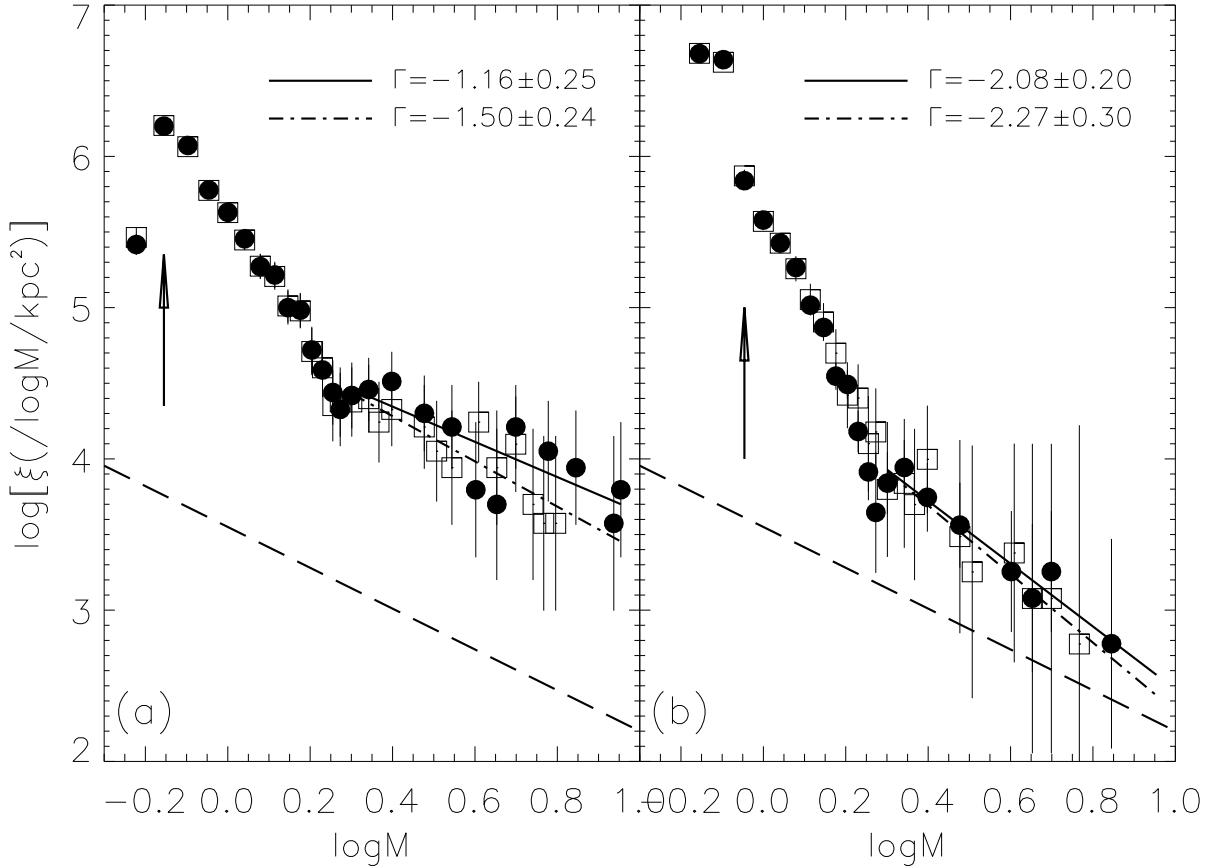,width=17.truecm,angle=0}
}}
\caption{The main-sequence MF of (a) the area on LH 52 and (b) the {\em
``field''} constructed by using the mass-luminosity relation provided by
two different isochrones (for $\log{\tau} \simeq$ 6.6 and 7.8).  It is
shown that for the low-mass stars the slopes of the MF are almost
identical, while for the intermediate-mass stars they are comparable. We
overplot the linear fits for every mass-luminosity relation for the
intermediate-mass range. The stars have been counted in logarithmic (base
ten) mass intervals. The numbers have been corrected for incompleteness
and normalized to a surface of 1 kpc$^{2}$.  The errors reflect the
Poisson statistics. The arrows indicate the 50\% completeness level. 
An offset Salpeter IMF has been overplotted with a dashed line for
reference.}
\label{imfiso}
\end{figure*}

One may consider that, with the use of a mass-luminosity relation for the
construction of the MF, uncertainties may occur if there is an age
distribution among the stars in the areas of interest. Then the use of a
single isochrone may not be adequate for a mass-luminosity conversion. We
tested this argument by constructing the MF of both fields with the use of
two different indicative isochrones, representative of the MS populations
observed in the fields (for $\log{\tau} \simeq$ 6.6 and 7.8). The first
choice is based on the fact that the area of LH 52 includes a very young
population, as shown in the CMDs of Fig. \ref{lh52cmdiso}, while the
second is taken as the younger limit of the older population observed in
both the CMDs of Fig.  \ref{lh52cmdiso}, taking into account the fact that
the transformation of luminosities to masses does not differ at all with
the use of older models for main-sequence stars. The constructed MF is
shown in Fig. \ref{imfiso}.  As shown earlier, a single-age scenario for
both regions is not justified.  Still, the use of a single model is a
simplification justified by the fact that the correspondence of
luminosities to masses is not age sensitive for the main sequence,
especially for the fainter stars. Indeed, we found that for the low-mass
stars the slopes of the MF are almost identical for both models, while for
the intermediate-mass stars ($M$ \gapprox\ 2 M{\solar})  they are very
close to each other, within their uncertainties.  In Figs. \ref{imfs} and
\ref{imfiso} the arrows indicate the 50\% completeness limits for each
sample.

Taking under consideration that the two areas were observed within
different programs they do not have the same exposure times in the common
bands and thus, as shown earlier, their completeness functions are
different.  Consequently for the comparison of the MF slopes we cannot use
the whole mass range for the LH 52 area, which has better completeness.
This is also applicable to the field subtraction described later. The MF
of the LH 52 area was constructed for masses down to $\sim$ 0.6 M{\solar}
and it is complete (50\% completeness) down to the limit of $\sim$
0.7 M{\solar}, while the area of the {\em ``field''} reaches the 50\%
completeness limit at $\sim$ 0.9 M{\solar}. We present the MF slopes of
the areas for the mass ranges with $M$ \lapprox\ 2 M{\solar} (low-mass
range) and $M$ \gapprox\ 2 M{\solar} (intermediate-mass range) in Table 2.
In this table we also present the slopes as they are found for the whole
of the complete mass range of each sample and we give the estimated slopes
of the MF, as it is constructed with the use of the mass-luminosity
relations derived from both isochrone models.

%\clearpage
%%%%%%%%%%%%%%%%%%%%%%%%%%%% FIGURE %%%%%%%%%%%%%%%%%%%%%%%%%%%%%%%%%%%%%%%
%%%%%%%%%%%%%%%%%%%%%%%%%%%%%%%%%%%%%%%%%%%%%%%%%%%%%%%%%%%%%%%%%%%%%%%%%%%
%\clearpage

From the values of Table 2 it can be seen that the MF of the {\em
``field''} is {\em systematically steeper} than the one of the LH 52 area
for the same mass ranges.  This finding is in line with the results by
Massey and collaborators (see e.g. Massey 1990; 1993), who found that the
MF slope of the {\em massive stars} in the general LMC field is steeper
than the slope of the OB associations. It was also found to be comparable
to the one of our Galaxy. The slopes given in Table 2 are comparable
to the PDMF slopes derived for the Galactic field by Chabrier (2003) for
$M >$ 1 M{\solar} (his table 1). Chabrier (2003) has also found a trend of
the MF slope to become shallower for higher-mass ranges in line with our
results. The fact that the general LMC field, which is represented by our
{\em ``field''} has indeed a steeper MF slope than the one of the LH 52
field is possibly connected to the fact that the latter includes an
association. This difference in the MF slope between the LH 52 area and
the {\em ``field''} is verified for every selected mass range shown in
Table 2. A reasonable explanation is the fact that the {\em ``field''}
seems to cover a large number of low-mass stars, while the area of LH 52
has a definite larger number of more massive MS stars, as it is discussed
earlier for the CMDs of the fields. Indeed, as it is shown in Figs.
\ref{imfs} and \ref{imfiso} the low-mass range of the MF up to about 1.25
M{\solar} is covered mostly by the {\em ``field''}, while the LH 52 area
has more highly populated bins of larger masses, which leads to a
top-heavy MF as discussed earlier.

A general conclusion based on the slopes in Table 2 is that in all cases,
except for the intermediate-mass range in the LH 52 area, the MF {\em does
not follow the Salpeter law, but being steeper}. Massey et al.  (1995)
found that the slope of the IMF of the LMC field stars is very steep,
$\Gamma = -4.1 \pm 0.2$, for stars with masses \gapprox\ 2 M{\solar}. This
slope is in very good agreement with {\em ``field''} $\Gamma$ for its
whole mass range ($\sim$ 1 - 7 M{\solar}). If we assume that there is no
change of the MF slope at masses around 2 M{\solar}, this slope for the
LMC field stars stays unchanged down to 1 M{\solar}. On the other hand
assuming that indeed the MF slope changes at $\sim$ 2 M{\solar}, then we
see that the MF of the LMC general field is very steep at the low-mass
range, it becomes more shallow (close to the Salpeter's) at the
intermediate-mass range, and becomes steep again for the massive stars.
Our results seem to favor this scheme. It should be noted that, as
seen from the slopes of Table 2, the intermediate-mass range MF slope for
the LH 52 area coincides with Salpeter's value, for all three selected
mass ranges and for both mass-luminosity relations.

So far we use the term ``Mass Function'' to describe the observed
distribution of stellar masses in both WFPC2 fields. The reason is that
the general LMC field, which dominates both regions, is found to include
stars old enough for intermediate-mass stars to have evolved from the main
sequence. On the contrary, the lower-mass main-sequence stars, which we
detect in our CMDs are expected to remain in their initial positions until
today. Under these circumstances the constructed mass distributions for
both areas represent their PDMF for their entire mass range, while for the
low-mass stars we can refer to it as the IMF.

%\clearpage
%%%%%%%%%%%%%%%%%%%%%%%%%%%% TABLE 2 %%%%%%%%%%%%%%%%%%%%%%%%%%%%%%%%%%%%%%%%%%
%\setcounter{table}{1}  
%%%%%%%%%%%%%%%%%%%%%%%%%%%%%%%%%% TABLE %%%%%%%%%%%%%%%%%%%%%%%%%%%%%%%%%%%%%%%%%%%%%%%%%%
\begin{table*}
%\begin{sidewaystable}
\begin{center}
\caption{Slopes of the main-sequence MF of the LH 52 area and the {\em ``field''}
for various mass ranges. \label{tab2}}
\begin{tabular}{c|cc|cc} 
\hline 
\hline
Mass Range  &\multicolumn{2}{c}{LH 52 area} & \multicolumn{2}{c}{\em ``field''} \\
(M{\solar}) &$\log{\tau}=6.6$&$\log{\tau}=7.8$&$\log{\tau}=6.6$&$\log{\tau}=7.8$\\
\hline 
\multicolumn{5}{c}{\em Small Masses} \\
\hline 
0.7 - 2.0  & $-$4.35 $\pm$ 0.15 & $-$4.37 $\pm$ 0.16 & - & - \\
0.9 - 2.0  & $-$4.51 $\pm$ 0.20 & $-$4.53 $\pm$ 0.22 & $-$6.57 $\pm$ 0.40 & $-$5.54 $\pm$ 0.21\\
%1.0 - 2.0  & $-$4.69 $\pm$ 0.23 & $-$4.71 $\pm$ 0.26 & $-$6.92 $\pm$ 0.46 & $-$5.60 $\pm$ 0.27\\
\hline 
\multicolumn{5}{c}{\em Intermediate Masses} \\
\hline 
2.0 - 6.0  & $-$1.24 $\pm$ 0.46 & $-$1.50 $\pm$ 0.24 & $-$2.03 $\pm$ 0.28 & $-$2.27 $\pm$ 0.31\\
2.0 - 7.0  & $-$1.11 $\pm$ 0.37 & - & $-$2.08 $\pm$ 0.20 & - \\
2.0 - 9.0  & $-$1.16 $\pm$ 0.26 & - & - & -\\
\hline 
\multicolumn{5}{c}{\em Whole Mass Range} \\
\hline 
0.7 - 6.0  & $-$2.48 $\pm$ 0.23 & $-$2.54 $\pm$ 0.18 & - & - \\ 
0.9 - 6.0  & $-$2.20 $\pm$ 0.24 & $-$2.33 $\pm$ 0.20 & $-$3.66 $\pm$ 0.35 & $-$3.69 $\pm$ 0.23\\
%1.0 - 6.0  & $-$2.07 $\pm$ 0.25 & $-$2.21 $\pm$ 0.20 & $-$3.49 $\pm$ 0.36 & $-$3.56 $\pm$ 0.24\\
0.7 - 7.0  & $-$2.31 $\pm$ 0.22 & - & - & - \\
0.9 - 7.0  & $-$2.03 $\pm$ 0.23 & - & $-$3.39 $\pm$ 0.31 & - \\
%1.0 - 7.0  & $-$1.91 $\pm$ 0.23 & - & $-$3.23 $\pm$ 0.31 & - \\ 
0.7 - 9.0  & $-$2.11 $\pm$ 0.19 & - & - & - \\
0.9 - 9.0  & $-$1.87 $\pm$ 0.18 & - & - & - \\
%1.0 - 9.0  & $-$1.77 $\pm$ 0.19 & - & - & - \\
\hline 
\hline
\end{tabular}
\end{center}
\tablecomments{The slopes estimated with the use of
mass-luminosity relations derived from two different models (for
$\log{\tau} \simeq$ 6.6 and 7.8) are given for the sake of comparison.
Symbol ``-'' marks the cases, for which the MF slope could not be estimated
due to the lack of stars in the specified mass ranges.}
%\end{sidewaystable}
\end{table*}
%%%%%%%%%%%%%%%%%%%%%%%%%%%%%%%%%%%%%%%%%%%%%%%%%%%%%%%%%%%%%%%%%%%%%%%%%%%%%%%%%%%%%%%%%

%%%%%%%%%%%%%%%%%%%%%%%%%%%%%%%%%%%%%%%%%%%%%%%%%%%%%%%%%%%%%%%%%%%%%%%%%%%%%%%
%\clearpage

\subsection{The Initial Mass Function of the Association LH 52}

\begin{figure*}[t!]
\centerline{\hbox{
\psfig{figure=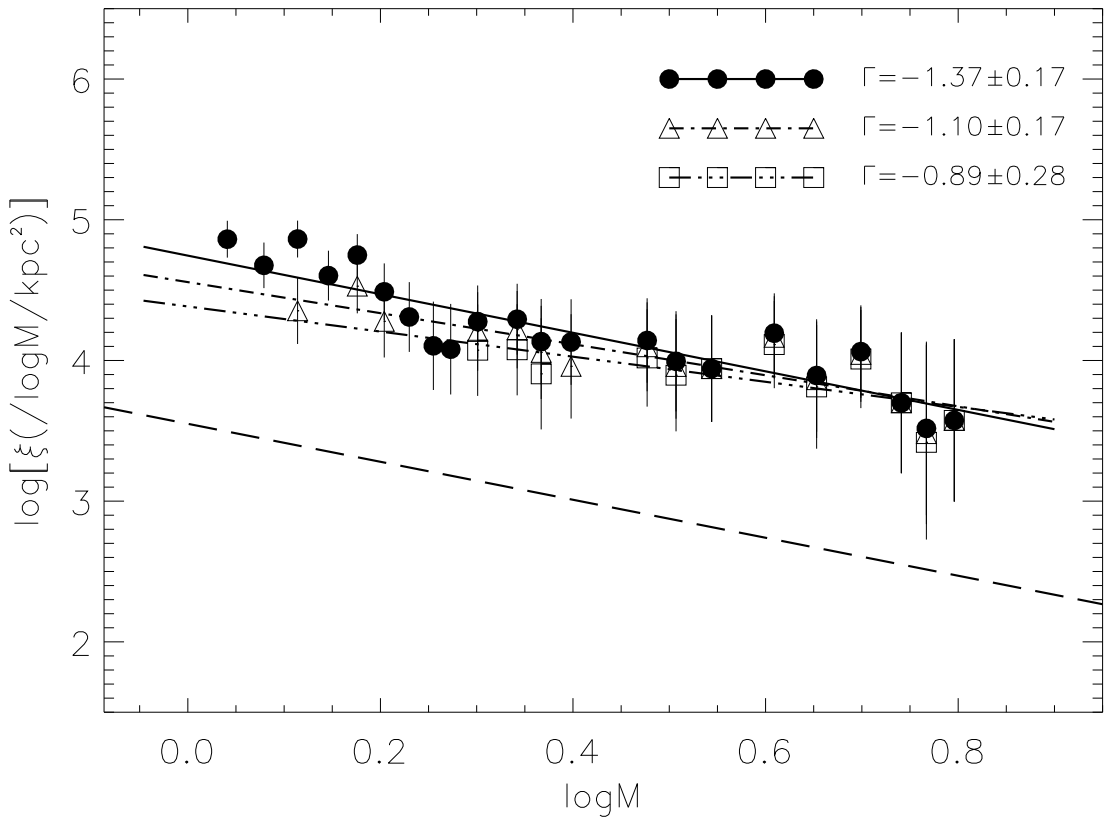,width=17.truecm,angle=0}
}}
\caption{The completeness corrected IMF of the southern part of
association LH 52 covered by the first of our fields after the general LMC
field contribution has been subtracted, assuming that it is well
represented by our second WFPC2 field. Three characteristic slopes,
estimated for the IMF constructed with three different assumptions on the
contribution of the general LMC field, have been overplotted. These slopes
are comparable with a somewhat shallower-than-Salpeter slope in the mean, 
although a Salpeter-type slope cannot be excluded in view of the observational
uncertainties. An offset Salpeter IMF has been overplotted with a dashed
line for reference.}
\label{imffin}
\end{figure*}

The MF of the LH 52 area is affected by the presence of the stellar
population of the general field of the galaxy. This was shown earlier in
the CMD of the area, where an underpopulated sub-giant branch, which seems
to be mixed with the population belonging to the association, points to a
contribution from the general LMC field. In order to disentangle this
contribution and to construct the MF of the stellar system in the LH 52
area (the association itself) we use the observations of the {\em
``field''}, which we treat as a representative sample of the general LMC
field.  Thus, we subtract the contributing number of {\em ``field''} stars
per mass bin from the corresponding in the LH 52 area. The remaining
numbers are used for the construction of the MF of the association.
Specifically, the number of low-mass main-sequence stars in different
mass ranges (taking completeness into account) of both CMDs is being used
to normalize the MF of the {\em ``field''} to the expected distribution if
the LH 52 area would contain only the field population. The derived
completeness corrected, field subtracted MF of the association LH 52 is
shown in Fig. \ref{imffin}. Since the system is found to be very
young this MF is actually the IMF (see also \S 5.1) of the
southwestern part of the association LH 52, which is covered by the first
HST/WFPC2 field presented here. In Fig. \ref{imffin} we plot the IMF,
as it is constructed after the field subtraction, assuming three different
ratios of the number of faint main-sequence stars in the {\em ``field"}
over the area on LH 52 for the estimation of the contribution of the
general LMC field.

Field subtraction is one of the major uncertainties in the determination
of the MF of a stellar system.  In \S 4.2 we note that the mean percentage
of the stars detected in the {\em ``field''} that has to be used for the
field subtraction is 58\% ($\pm$ 7\%). We performed this subtraction
considering the two extremes of the expected range of the field
contribution (37\% and 91\% of the stars in the {\em ``field''}) in order
to estimate the uncertainty in the slope of the IMF of LH 52 due to this
subtraction. The corresponding IMFs have been overplotted with different
symbols in Fig. \ref{imffin}, and their slopes derived from the
corresponding linear fits, are given.  All linear fits applied for the
evaluation of the IMF slope are not weighted. It should be noted that, as
shown in Fig.  \ref{imffin}, larger percentage of the contribution of the
field to the IMF of the association leads to fewer low-mass bins. This is
a reasonable result, since, as shown earlier, the low-mass population in
both observed fields seems to belong mostly to the general background
field. The field subtraction is valid for masses larger than 0.9
M{\solar}, since this is the mass limit, where the stellar sample of the
{\em ``field''} is complete (50\% completeness). Consequently we are
able to estimate the IMF slope of the system for masses $M$ \gapprox\ 1
M{\solar} (or larger, depending on the assumption for the the field
subtraction).

%\clearpage
%%%%%%%%%%%%%%%%%%%%%%%%%%%% FIGURE %%%%%%%%%%%%%%%%%%%%%%%%%%%%%%%%%%%%%%%
%%%%%%%%%%%%%%%%%%%%%%%%%%%%%%%%%%%%%%%%%%%%%%%%%%%%%%%%%%%%%%%%%%%%%%%%%%%
%\clearpage

The derived slope of the association was found to be comparable to a
Salpeter IMF, but with a trend to be shallower, depending on the assumed
field contribution. Specifically its value is found to lay between the
limits of $\simeq$ $-$0.90 and $-$1.37 (Fig. \ref{imffin}), which result
in a IMF slope for LH 52 $\Gamma \simeq -1.12 \pm 0.14$ for the mass range
given above.  This slope is in line with the average value of the IMF
slopes of the associations in the Milky Way ($\Gamma \simeq -1.1 \pm
0.1$), but somewhat more shallow than the one of the Magellanic Clouds
($\Gamma \simeq -1.3 \pm 0.1$), found by Massey, Johnson \&
Degioia-Eastwood (1995) for stars with masses $>$ 7 M{\solar}. It is
also in agreement with the global Galactic disk IMF slope as derived by
Chabrier (2003) for $M >$ 1 M{\solar}. These facts suggest that the IMF of
stellar associations has more or less the same slope for intermediate- and
high-mass stars. From Fig. \ref{imffin} it can be seen that, after the
field subtraction, the IMF slope seems to be constant through the whole
mass range and does not change significantly for different mass ranges.  
However we estimated the IMF slope of the association for the two mass
ranges separated by the limit of 2 M{\solar}, for reasons of comparison
with the results in the observed fields. We found that if the mean
contribution by the field is assumed ($\sim$ 58\%) then the IMF slope for
the low- ($\sim$ 1 - 2 M{\solar}) and for the intermediate-mass range
($\sim$ 2 - 7 M{\solar}) remains unchanged with a value between $-$1.07
and $-$1.08. On the other hand, if a modest contribution by the field is
assumed ($\sim$ 37\%), then the low-mass IMF has a steep slope of $\simeq
-3.1 \pm 0.3$, much different from the one for the intermediate-mass
range, which was found $\simeq -1.3 \pm 0.2$. If a high contribution by
the field is assumed ($\sim$ 91\%), then we can only estimate the IMF
slope for the intermediate-mass range (M \gapprox\ 2 M{\solar}), which is
the one given in Fig. \ref{imffin} ($-0.9 \pm 0.3$). From these values
one can conclude that the stars with masses larger than 2 M{\solar}
determine the total IMF slope of the association. This is in line with our
results so far, which imply that the general LMC field is responsible for
the low-mass population observed in both fields, while the MF of the LH 52
area is mostly affected by the existence of more massive stars, members of
the association. Under these circumstances there are clear indications
that the IMF of a star forming stellar association in the LMC is a
``top-heavy'' IMF, due to its intermediate-mass stellar members, but the
uncertainties are too large for this to be a definite conclusion.

\section{Summary and Discussion}

In this paper we studied the MF, using HST/WFPC2 observations, of two
stellar fields located on the boundary of the super-shell LMC 4, which is
surrounded by a large number of star-forming stellar associations. The
first field targets the southwestern part of the known large association
LH 52, which resides (with LH 53) on the border of LMC 4 with the shell
LMC 5 (Fig. \ref{lmc4map}). The second field is located in the empty area
between two smaller associations (LH 54 and LH 55) towards the south of
LMC 4 (we refer to it as the {\em ``field''})  and it accounts for the
general background field of the LMC, which deserves an equally thorough
investigation. Both areas are populated by a large number of faint stars
with masses down to the sub-solar regime. The CMD of the LH 52 field,
though, reach a much brighter upper MS limit, while for the {\em
``field''} there is a lack of MS stars brighter than $V \simeq$ 18 mag.
Its brighter population is mostly characterized by a prominent sub-giant
branch.

We construct the MF of the stellar populations detected in both fields,
taking into account several considerations concerning the sources of
uncertainty in the derived slopes. The MF is constructed with two methods
for counting stars in mass intervals. For the first we use a
mass-luminosity relation for transforming luminosities into masses and for
the second we transform colors and magnitudes into temperatures and
luminosities, and we count stars between evolutionary tracks according to
their positions in the HR-Diagram. We verify, as it was previously found,
that both methods give comparable results concerning the slope of the
constructed MF. As far as the mass-luminosity relation concerns, we test
the possibility that the use of theoretical models of different ages may
result in the construction of much different MF, and in consequence may
give different results for the corresponding slope. We found that as far
as the MS population concerns, there are no significant changes in the
slopes of the constructed MF, though the upper mass limit depends on the
selected mass-luminosity relation.

The large majority of the observed low-mass stars with 1 \lapprox\ $M$
\lapprox\ 2 M{\solar} in both fields seems to be a feature of the general
field of the LMC. The MF of these stars resembles very well their IMF,
since they are not expected to have left the MS, within the life-span of
the LMC general field, as it is revealed from the CMD of the {\em
``field"} (\S 4.3). The slope of this MF is found to be very steep,
between $\Gamma= -4$ and $-6$ and {\em systematically} much steeper than
the one of the intermediate-mass range (2 \lapprox\ $M$ \lapprox\ 7
M{\solar}). This implies that the contribution of the low-mass MS stars to
the MF is higher than expected, if it is assumed that the MF follows the
same slope for masses smaller than $\sim$ 2 M{\solar}. The general LMC
field has an intermediate-mass MF slope around $\Gamma= -2$, which is
steeper than the corresponding one of the LH 52 area. The latter is found
to be very close to Salpeter's value with $\Gamma$ between $-1$ and $-1.5$
(Table 2), comparable to the typical MF slope previously found for the
massive stars in associations of the LMC ($\Gamma \simeq -1.4$;  Massey et
al. 1995). The MF slope for the whole mass range (1 \lapprox\ $M$
\lapprox\ 7 M{\solar}) in both fields falls between the above limits, with
the one of the {\em ``field''} being definitely steeper than the one on
the association LH 52.

The slope of the field MF was found previously to be steeper (of the order
of $\Gamma \simeq -3$ to $-4$) than of the stellar associations in the LMC
and the Galaxy, for the most massive stars with $M$ \gapprox\ 7 M{\solar}
(Massey, Johnson \& Degioia-Eastwood 1995). This was verified for stars
with smaller masses down to 2 M{\solar} more recently by Gouliermis et al.
(2002), who investigated photometrically the stellar content and the MF of
three stellar associations and their fields located on the edges of LMC 4.
These authors found that the MF slope of the association LH 95 for
intermediate-mass stars is somewhat steeper than Salpeter's (1955) with
$-1.9 \leq \Gamma \leq -1.2$ and, indeed, more shallow than in the
surrounding general field of LMC ($-4.9 \leq \Gamma \leq -3.6$).  The
latter result is further supported by the larger number statistics of our
investigation here. We verify that {\em the MF of the general field of the
LMC is steeper than the MF of the area, where a stellar association is
located and of the IMF of the association, corrected for the field
contribution, also toward lower masses down to $M \sim$ 1 M{\solar}}.

We use the MF of the {\em ``field''} for correcting the field
contamination in the observed area of the association LH 52. Thus, we
construct the field-subtracted, incompleteness-corrected, main-sequence MF
of the southwestern part of the association itself. This MF is
actually the IMF of the system, since it is considered to be very young.  
Its construction was possible for masses down to about 1 M{\solar}, due
to the completeness limitations in the { \em ``field''}. We find that the
slope of the IMF of the association is {\em comparable to, but more
shallow than a typical Salpeter IMF}. This is due to the large number of
more massive stars in the area, which implies that this IMF may be
considered as a top-heavy IMF. This result is in agreement with the one by
Hill et al. (1994), according to which {\em the shallower slope of the
association IMF suggests that not only the star formation rate is higher
in associations, but the local conditions favor the formation of higher
mass stars}.  The IMF slope of LH 52 is found here to be $\Gamma
\simeq -1.4 \pm 0.2$ for masses M \gapprox\ 1 M{\solar} (if $\sim$ 37\% of
the {\em ``field''} is assumed to contaminate the population of LH 52),
$\Gamma \simeq -1.1 \pm 0.2$ for masses M \gapprox\ 1.5 M{\solar}
(assuming that $\sim$ 58\% of the {\em ``field''} population contributes
to the area on LH 52), and $\Gamma \simeq -0.9 \pm 0.3$ for a shorter mass
range (M \gapprox\ 2 M{\solar}) (if the field contamination of the LH 52
area corresponds to $\sim$ 91\% of the stars in the {\em ``field''}).  
These values are very close to what previously was found by HCB95 for the
same association ($\Gamma \simeq -1$, for $M$ \gapprox\ 3.5 M{\solar}).

The MF of the intermediate-mass stars observed by Gouliermis et al. (2002)  
in the area of the association LH 95 (including the contribution of the
LMC field)  was found with a slope $-1.8 \leq \Gamma \leq -1.5$ comparable
to our results on both observed fields in the same mass range (Table 2).
These values agree with the slopes found for 14 associations in the
Magellanic Clouds by Hill, Madore \& Freedman (1994; $\Gamma \simeq -2.0
\pm 0.5$ for $M >$ 9 M{\solar}). The field subtracted IMF of the
associations LH 95 and LH 52 themselves, though, are more shallow and in
line with the ones presented previously by Massey et al. (1995; $\Gamma
\simeq -1.3 \pm 0.3$). The latter authors note that the differences in the
estimated IMF slopes between their study and the one by Hill et al. is due
to {\em the inability of photometry alone (with no spectroscopy) to treat
the hottest (and most massive) stars}. Here we stress the importance of
the field contamination of the IMF of an association, especially toward
smaller masses. The difference in the IMF slope due to the field
contribution, was already noted by Parker et al. (2001) for the
association LH 2. These authors state that {\em ``it is possible that even
in the case of an universal IMF, the variability of the density of field
stars may play an important role in creating the observed differences
between calculated IMFs for OB associations''}. Our results support this
suggestion.

Concerning the universality of the IMF, Hunter (1995) has tabulated IMF
slopes and their appropriate stellar mass ranges for stellar systems in
the Local Group, showing that the slope of the IMF of stellar associations
is lying within the same limits of $-$1.0 $\pm$ 0.1 and $-$1.7 $\pm$ 0.2.  
The IMF of these studies has been determined for stars more massive than
25 M{\solar} only, and for stars with masses larger than $\sim$ 7
M{\solar}.  Hunter et al. (1997) note that no statistically significant
variation can be seen among the slopes, but the uncertainties are fairly
high. They conclude that for young star clusters and associations the IMF
is independent of the galactic characteristics and the mass range and that
the star formation process is truly local, independent of the galaxy in
which it is located. Sirianni et al.  (2000) studied the low-end IMF of R
136 and they found an agreement of their IMF slope with earlier
investigations for masses larger that $\sim$ 3 M{\solar}, but they report
that the IMF shows a definite flattening below $\simeq$ 2 M{\solar}.
According to these authors, these findings add to the conclusion of Scalo
(1998) that, {\em at least for stars less massive than $\sim$ 1 M{\solar},
the IMF is not uniform}. A flattening of the IMF could also be indicated
for masses below 1 M{\solar} by Sirianni et al. (2002) in HST/WFPC2
observations of the cluster NGC 330 in the SMC. Yet, as these authors
state, ``since the photometric completeness in that mass range is always
less than 50\%, one cannot assess the reliability of this intriguing
conclusion, which would require deeper photometry''.

Our investigation shows that, as far as the general field of the LMC
concerns, {\em the MF slope depends on the mass range}, being very steep
for stars with masses smaller than $\sim$ 2 M{\solar}, and very much
different for larger masses (Figs. \ref{imfs} \& \ref{imfiso}). On the
other hand the IMF of the star-forming stellar association, as seen in
Fig. \ref{imffin} after the field subtraction, seems to follow the same
slope for the whole observed mass range. So, within the 50\% completeness
of our data, {\em we did not identify any flattening of the IMF for stars
with $M$ \lapprox\ 2 M{\solar}}. This result clearly suggests that {\em
the IMF of a young stellar system in the LMC may be considered to be the 
same through the whole of its detected mass range}. In addition, in
our data we could not identify any lower mass cutoff in the IMF for the
detected MS stars, in line with the results by Brandner et al. (2001), who
found no evidence for a lower mass cutoff in the IMF for pre-main sequence
stars in the vicinity of 30 Doradus in the LMC.

Some final remarks that can be additionally drawn from the study presented
here are: 1) HST observations with WFPC2 on associations in the LMC reveal
their sub-solar stellar masses with reasonable completeness, due to lack
of crowding. 2) In consequence the IMF of these systems can be
successfully constructed down to the detected low-mass end, in the
optical, due to low reddening. 3) Stellar associations cover length-scales
larger than typical star clusters, and thus, wide field imaging is
required for the study of a typical extra-galactic stellar association at
its extend. Hence, the {\em Advanced Camera for Surveys} would be the most
appropriate instrument for such studies to be carried on.

\acknowledgments

This paper is based on observations made with the NASA/ESA Hubble Space
Telescope, obtained from the data archive at the Space Telescope Science
Institute. STScI is operated by the Association of Universities for
Research in Astronomy, Inc. under NASA contract NAS 5-26555.

%%%%%%%%%%%%%%%%%%%%%%%%%%%  BIBLIOGRAPHY  %%%%%%%%%%%%%%%%%%%%%%%%%%%%%%%
%%%%%%%%%%%%%%%%%%%%%%%%%%%%%%%%%%%%%%%%%%%%%%%%%%%%%%%%%%%%%%%%%%%%%%%%%%
%\newpage

%%%%%%%%%%%%%%%%%%%%%%%%%%%%%%%%%%%%%%%%%%%%%%%%%%%%%%%%%%%%%%%%%%%%%%%%%%

\end{document}